\documentclass[11pt,a4paper]{JHEP3}
\preprint{MITP/14-024}

\pdfoutput=1

\usepackage{snapshot}
\usepackage{cite}
\usepackage{amssymb}
\usepackage{amsmath}
\usepackage{epsfig}
\usepackage{subfigure}
\usepackage{color}
\let\normalcolor\relax
\usepackage{graphicx}
\usepackage{inputenc}
\usepackage{xspace}
\usepackage{slashed}
\inputencoding{latin1}
\usepackage{axodraw4j}
\usepackage{array}

\def\lsim{\mathrel{\raise.3ex\hbox{$<$\kern-.75em\lower1ex\hbox{$\sim$}}}}
\def\gsim{\mathrel{\raise.3ex\hbox{$>$\kern-.75em\lower1ex\hbox{$\sim$}}}}
\def\ifmath#1{\relax\ifmmode #1\else $#1$\fi}

\newcommand{\be}{\begin{equation}}
\newcommand{\ee}{\end{equation}}
\newcommand{\bea}{\begin{eqnarray}}
\newcommand{\eea}{\end{eqnarray}}
\title{\boldmath{Fine Tuning in the Holographic Minimal Composite Higgs Model}}

\author{Paul R. Archer
\\ PRISMA Cluster of Excellence \& Mainz Institute for Theoretical Physics Johannes Gutenberg University, 55099 Mainz, Germany.
}

\abstract{In the minimal composite Higgs model (MCHM), the size of the Higgs mass and vacuum expectation value is determined, via the Higgs potential, by the size of operators that violate the global SO(5) symmetry. In 5D holographic realisations of this model, this translates into the inclusion of brane localised operators. However, the inclusion of all such operators results in a large and under-constrained parameter space. In this paper we study the level of fine-tuning involved in such a parameter space, focusing on the MCHM${}_5$. It is demonstrated that the gauge contribution to the Higgs potential can be suppressed by brane localised kinetic terms, but this is correlated with an enhancement to the S parameter. The fermion contribution, on the other hand, can be enhanced or suppressed. However this does not significantly improve the level of fine tunings, since the Higgs squared term, in the potential, requires a cancellation between the fermion and gauge contributions. Although we focus on the MCHM${}_5$, the fermion holographic Lagrangian - including all possible brane localised mass and kinetic terms - is derived in the appendix and will have applications to a wider range of composite Higgs models. }


\begin{document}
\section {Introduction}
\label{sec:intro}

The discovery of the Higgs boson, in addition to observational evidence for its scalar nature \cite{Aad:2013xqa, Chatrchyan:2013mxa}, as well as the discovery of no other deviation from the standard model (SM), has forced particle physics into an uncomfortable position. Although, it should be remembered that the Higgs is not the first scalar that has been discovered to have a mass lower than the scale of physics associated with its production. In particular, no one speaks of a hierarchy problem associated with the pion, since its mass ($\sim140$ MeV) is only slightly lower than the QCD scale ($\sim 200$ MeV), which is explained by the logarithmic running of $\alpha_s$. It was perhaps such considerations that led to the proposal that the relative lightness of the Higgs can be explained if it is a pseudo-Nambu-Goldstone boson (pNBG) of a strongly coupled sector \cite{Terazawa:1976xx, Terazawa:1979pj, Kaplan:1983fs, Kaplan:1983sm, Dimopoulos:1981xc}. The Higgs mass is then protected by an approximate global symmetry, analogous to chiral symmetry in QCD.  

Nonetheless, there are two principal differences between a pNBG Higgs and the pion. Firstly, the Higgs appears to be considerably lighter than the scale of new physics. Secondly, early measurements of the Higgs couplings indicate that the fermions and W/Z Bosons are predominantly gaining their mass from the vacuum expectation value (VEV) of the Higgs, as in the SM. Hence, the Higgs must have an effective potential such that it gains a non-zero VEV, i.e. it must have a negative Higgs squared term. The relevant question is then how `natural' is this or equivalently how much fine tuning is required in order for this to happen?

Of central importance, to addressing this question, is being able to calculate the coefficients of the Higgs potential, generated via the Coleman-Weinberg mechanism \cite{Coleman:1973jx}. However, one is working with a strongly coupled theory and so it is necessary to either work with lattice simulations or find an effective theory valid to scales, such that the potential can still be reliably estimated. Progress, with this latter option, was made by using the AdS / CFT conjecture to relate this scenario to a 5D model of gauge-Higgs unification (see \cite{Agashe:2004rs, Hosotani:1983xw, Hosotani:1983vn, Manton:1979kb, RandjbarDaemi:1982hi, Haba:2004qf, Hosotani:2005nz, Sakamura:2007qz,  Hosotani:2008tx, Scrucca:2003ra, Contino:2003ve, Panico:2005dh, Falkowski:2006vi, Medina:2007hz} and references therein). In such models the Higgs potential (and hence the Higgs mass and VEV) are determined by the size of the operators that violate the global symmetry, i.e. the analogues of the Yukawa couplings in chiral symmetry breaking. In the language of the 5D model, these would correspond to brane-localised operators. So, for example, if you were to include no brane localised operators, then one would find that the Higgs was massless and there would be no electroweak symmetry breaking (EWSB).

Previous studies, for simplicity, have often just included brane localised mass terms. However, the 5D theories are  effective theories and hence, unless the precise dynamics responsible for breaking the global symmetry are specified, one should include all operators permitted by the unbroken symmetry. If not initially included, such operators would be generated by radiative corrections to the 5D theory \cite{Georgi:2000ks}. With this in mind, the purpose of this paper is twofold. Firstly, we wish to make a quantitative study of how much fine tuning is required for the minimal composite Higgs model (MCHM) \cite{Agashe:2004rs} and secondly to investigate the effect of, in particular, brane localised kinetic terms on the Higgs potential.

A known problem with the MCHM is there are a large number of model building possibilities, see \cite{Contino:2006qr, Contino:2006nn, DeCurtis:2011yx, Panico:2011pw, Carmona:2012jk, Barbieri:2012tu, Gillioz:2013pba, Carmona:2013cq, Carena:2014ria, Bellazzini:2014yua}. Here we restrict ourselves to studying an AdS${}_5$ geometry, which would correspond to studying a subclass of possible models with an approximate conformal scaling between the TeV scale and the Planck scale. We also restrict our study to a version of the MCHM with the minimal fermion content (that still avoids large corrections to $Z\to b_L\bar{b}_L$), notably the MCHM${}_5$ \cite{Contino:2006qr}. As was explained in \cite{Panico:2012uw}, it is expected that the tuning required to achieve EWSB will be greater in the MCHM${}_5$ than in alternative models, including for example \textbf{14}'s \cite{Pappadopulo:2013vca}. However, it was found in \cite{Carena:2014ria} that the tuning required to obtain the correct Higgs mass can increase in such models. It is also important to quantifiably understand the tuning of the minimal models before one extends them.

In this paper we relegate the technical details of the calculation to the appendices, where we compute the holographic Lagrangian for the gauge fields (appendix \ref{App:Gauge}) and fermion fields (appendix \ref{App:Fermions}), before applying them to the MCHM in appendix \ref{App:MCHM}. The details of the MCHM are discussed in section \ref{sec:MCHM}, while the Higgs potential is examined in more detail in section \ref{sec:higgspot}. Finally our numerical results are presented in section \ref{sec:numerics} and we conclude in section \ref{sec:conclu}.      

\section{The Holographic Minimal Composite Higgs Model}
\label{sec:MCHM}
In this paper we will study an unspecified strongly coupled theory with a global $\mathrm{SO}(5)\times\mathrm{U}(1)_X$ symmetry. The theory is assumed to be approximately conformal between a UV cut-off, around the Planck scale, and an IR cut-off at the scale $M_{\mathrm{KK}}\sim\mathcal{O}(\mathrm{TeV})$. The IR cut-off will break the global symmetry $\mathrm{SO}(5)\times\mathrm{U}(1)_X$ to a global $\mathrm{SO}(4)\times\mathrm{U}(1)_X$, of which a $\mathrm{SU}(2)_L\times\mathrm{U}(1)_Y$ subgroup will be gauged to form the SM. A massless Higgs boson would then arise as the Nambu-Goldstone boson of this symmetry breaking. However, associated with this IR cut-off will be operators that violate the $\mathrm{SO}(5)$ symmetry and result in the Higgs being a massive pNGB, which would then gain a non-zero vacuum expectation value (VEV), as in the SM. 

We give details of the holographic description of the MCHM in appendix \ref{App:MCHM}, although here we shall emphasise a few main points. It is conjectured that the above scenario can be described by a 5D theory of gauge-Higgs unification \cite{Hosotani:1983xw, Hosotani:1983vn, Manton:1979kb, RandjbarDaemi:1982hi}. At the level of the effective Lagrangian, any 5D model of gauge-Higgs unification can be expressed, in the holographic basis, as a 4D theory with a non-linear sigma field. However, by focusing on an AdS${}_5$ geometry, one can appeal to the AdS / CFT correspondence and conjecture that there is an exact correspondence with a strongly coupled, broken, conformal field theory \cite{Rattazzi:2000hs, PerezVictoria:2001pa, ArkaniHamed:2000ds}.    

So we will consider a 5D $\mathrm{SO}(5)\times\mathrm{U}(1)_X$ gauge field propagating in a slice of AdS${}_5$,  
\begin{equation}
\label{eqn:AdSMetric}
ds^2=\frac{R^2}{r^2}\left (\eta^{\mu\nu}dx_\mu dx_\nu-dr^2\right )=G^{MN}dx_Mdx_N
\end{equation}
with $\eta^{\mu\nu}=\mathrm{diag}(1,-1,-1,-1)$. The space is cut-off by branes in both the IR and the UV, $R=r_{\mathrm{UV}}\leqslant r\leqslant r_{\mathrm{IR}}=R^{\prime}$. Such a geometry can be characterised by two parameters, the Kaluza-Klein (KK) scale $M_{\mathrm{KK}}=\frac{1}{R^{\prime}}\sim\mathcal{O}(\mathrm{TeV})$ and the warp factor $\Omega=\frac{R^{\prime}}{R}\sim 10^{15}$. It should be stressed that such a scenario is very much an effective theory, valid up to some `warped down' cut-off at which one looses perturbative control of the 5D theory. The size of this cut-off is dependant on the size of other parameters in the model, but by naive dimensional analysis, is estimated to be   $\sim \mathcal{O}(10M_{\mathrm{KK}})$ \cite{Csaki:2008zd}.  

Likewise, the 5D branes should arguably be viewed as a framework with which one can parameterise our ignorance of the physics responsible for breaking the $\mathrm{SO}(5)$ symmetry. Given the effective nature of the 5D theory, one should endeavour to include all brane localised operators permitted by the symmetries of the theory. So here we will include all possible gauge kinetic terms \cite{delAguila:2003bh}.   

The Higgs arises as fluctuations along the four $\mathrm{SO}(5)/\mathrm{SO}(4)$ broken directions and can be parameterised by the non-linear sigma field 
\begin{equation}
\label{SigmaField}
\Sigma(x^\mu)=\exp\left [i\frac{\sqrt{2}T_C^{\hat{a}}\;h_{\hat{a}}(x^\mu)}{f_\pi}\right ],
\end{equation}
where $T_C^{\hat{a}}$ are the four broken generators of $\mathrm{SO}(5)$ (\ref{eqn:SO5generators}). The analogue of the pion decay constant is then defined by the coupling of $\Sigma$ with the W boson, at zero momentum. In 5D theories, this can be found by equating the holographic Lagrangian with $\mathcal{L}_{\Sigma}=\frac{f_\pi^2}{2}|D_\mu \Sigma|^2$ \cite{Contino:2003ve, Agashe:2004rs}, resulting in
\begin{equation}
\label{eqn:5Dfpi}
f_\pi^2=\frac{4M_{\mathrm{KK}}^2}{g^2\left (\log(\Omega)+\frac{\theta_{\mathrm{IR}}+\zeta_{\mathrm{IR}}}{1+\zeta_{\mathrm{IR}}}+\frac{\theta_{\mathrm{UV}}+\zeta_{\mathrm{UV}}}{1-\zeta_{\mathrm{UV}}}\right )},
\end{equation}   
where $g$ is the 4D gauge coupling of the SM. Due to an effective potential (to be computed in section \ref{sec:higgspot}), the Higgs will gain a non zero VEV in the direction of $T_C^{\hat{4}}$, i.e. $\langle h^{\hat{a}}\rangle=(0,0,0,h)$. After expanding the 5D action (\ref{eqn:5dGaugeAct}) in the holographic basis, the gauge sector is then described by \cite{Marzocca:2012zn},
\begin{align}
\label{eqn:LagPostEWSB}
    \textstyle{\mathcal{L}_{\mathrm{Hol.}}}&\textstyle{=-\frac{P_t^{\mu\nu}}{2}\;\bigg [\frac{2}{g_5^2}W_\mu^+\left (\Pi^{(+)}+\frac{1}{2}s_h^2\left (\Pi^{(-)}-\Pi^{(+)}\right )\right )W_\nu^-+A_\mu\left ( \frac{2s_w^2}{g_5^2}\Pi^{(+)}+\frac{c_w^2-s_w^2}{g_{5,X}^2}\Pi_X^{(+)}\right )A_\nu}\nonumber\\
    &\hspace{3cm}\textstyle{+Z_\mu \left (\frac{c_w^2+s_x^2s_w^2}{g_5^2}\Pi^{(+)}+\frac{c_x^2s_w^2}{g_{5,X}^2}\Pi_X^{(+)}+\frac{s_h^2}{2c_w^2g_5^2}\left (\Pi^{(-)}-\Pi^{(+)}\right )\right )Z_\nu}\nonumber\\
    &\hspace{6.5cm}\textstyle{+Z_\mu\;2c_ws_w\left (\frac{c_x^2}{g_5^2}\Pi^{(+)}-\frac{c_x^2}{g_{5,X}^2}\Pi_X^{(+)} \right )A_\nu\bigg ]},
\end{align}
where $s_w^2=1-c_w^2$ is the weak mixing angle, $s_x^2=1-c_x^2=s_w^2/c_w^2$ and $g_{5}$ is the 5D coupling. After examining (\ref{eqn:LagPostEWSB}), one finds that in order to obtain the correct W and Z masses, the Higgs VEV should be\footnote{In the literature, $s_h^2$ is also denoted by $\xi$ or $\epsilon$. It is sometimes taken to be a free parameter, however in holographic models its required value is uniquely fixed by the W mass and the geometry.}  
\begin{equation}
\label{eqn:sh2 }
\sin^2 \left (\frac{h}{f_\pi}\right )\equiv s_h^2=\frac{v^2}{f_\pi^2}=\frac{m_W^2}{M_{\mathrm{KK}}^2}\left (\log(\Omega)+\frac{\theta_{\mathrm{IR}}+\zeta_{\mathrm{IR}}}{1+\zeta_{\mathrm{IR}}}+\frac{\theta_{\mathrm{UV}}+\zeta_{\mathrm{UV}}}{1-\zeta_{\mathrm{UV}}}\right ),
\end{equation}
where the SM VEV is $v\approx 246$ GeV. Note that, as discussed in appendix \ref{App:Gauge}, in order to avoid tachyonic resonances it is required that
\begin{equation}
\label{eqn:gaugeTachCond}
\frac{\theta_{\mathrm{IR}}+\zeta_{\mathrm{IR}}}{1+\zeta_{\mathrm{IR}}}\geqslant 0\quad\mbox{ and }\quad \frac{\theta_{\mathrm{UV}}+\zeta_{\mathrm{UV}}}{1-\zeta_{\mathrm{UV}}}\geqslant 0.
\end{equation} 
Naively one would anticipate that $\theta_{\mathrm{IR},\;\mathrm{UV}}\sim \zeta_{\mathrm{IR},\;\mathrm{UV}}\sim \mathcal{O}(1)$ and hence the effects of such terms are expected to be subdominant to the $\log(\Omega)$ term.

\subsection{Electroweak Precision Tests}
\label{sect:EWPTs}

From the holographic Lagrangian (\ref{eqn:LagPostEWSB}), it is relatively straightforward to compute the leading order, oblique, contribution to the electroweak precision observables (EWPO's), traditionally parameterised in terms of the Peskin-Takeuchi parameters \cite{Peskin:1991sw}, 
\begin{align}
\frac{\alpha}{4s_w^2c_w^2}S&=\Pi^{\prime}_{ZZ}-\frac{c_w^2-s_w^2}{c_ws_w}\Pi^{\prime}_{Z\gamma}-\Pi^{\prime}_{\gamma\gamma},\nonumber\\
\alpha T&=\frac{\Pi_{WW}(0)}{m_W^2}-\frac{\Pi_{ZZ}(0)}{m_Z^2},\nonumber\\
\frac{\alpha}{4s_w^2}\left (S+U \right )&=\Pi^{\prime}_{WW}-\frac{c_w}{s_w}\Pi^{\prime}_{Z\gamma}-\Pi^{\prime}_{\gamma\gamma}, \label{eqn:STUdef}
\end{align} 
where $\Pi^{\prime}\equiv\partial_{p^2}\Pi(p) |_{p=0}$ and $\alpha$ is the fine structure constant. $\Pi^{\prime}_{ZZ}$, $\Pi^{\prime}_{WW}$, $\Pi^{\prime}_{Z\gamma}$ and $\Pi^{\prime}_{\gamma\gamma}$ can then be read off from (\ref{eqn:LagPostEWSB});
\begin{eqnarray}
1- \Pi^{\prime}_{ZZ}& = &\frac{g^2(c_w^2+s_x^2s_w^2)}{g_5^2}\Pi^{(+)\prime}+\frac{g^2c_x^2s_w^2}{g_{5,X}^2}\Pi_X^{(+)\prime}+\frac{g^2s_h^2}{2c_w^2g_5^2}\left (\Pi^{(-)\prime}-\Pi^{(+)\prime}\right ), \nonumber\\
1-\Pi^{\prime}_{\gamma\gamma}& = &\frac{2g^2s_w^2}{g_5^2}\Pi^{(+)\prime}+\frac{g^2(c_w^2-s_w^2)}{g_{5,X}^2}\Pi_X^{(+)\prime},\nonumber\\
1- \Pi^{\prime}_{WW}& = &\frac{g^2}{g_5^2}\Pi^{(+)\prime}+\frac{g^2s_h^2}{2g_5^2}\left (\Pi^{(-)\prime}-\Pi^{(+)\prime}\right ), \nonumber\\
  \Pi^{\prime}_{Z\gamma}&=&2c_ws_w\left (\frac{g^2c_x^2}{g_5^2}\Pi^{(+)\prime}-\frac{g^2c_x^2}{g_{5,X}^2}\Pi_X^{(+)\prime} \right ).
\end{eqnarray} 
By noting that $\Pi^{(+)}(p)/g_5^2\approx p^2/g^2+\mathcal{O}(p^4)$ and using (\ref{eqn:ApproxPiNeg}), it is found that the leading order contributions are
\begin{eqnarray}
S & = & \frac{8\pi s_h^2}{g_5^2}\left (\Pi^{(+)\prime}-\Pi^{(-)\prime}\right ) \nonumber\\
 & \approx &\frac{2\pi v^2}{M_{\mathrm{KK}}^2}\left (\frac{3}{4}+\frac{\theta_{\mathrm{IR}}+\zeta_{\mathrm{IR}}}{1+\zeta_{\mathrm{IR}}}\right ) +\mathcal{O}(\Omega^{-2}). \label{eqn:SParam} 
\end{eqnarray}    
Because of the custodial symmetry,
\begin{equation}
\label{Tparam }
 T=0\hspace{1cm}\mbox{ and }\hspace{1cm} U=0.
\end{equation}
Note, in accordance with \cite{Agashe:2007mc}, the $S$ parameter cannot become negative without giving rise to tachyonic modes. 

\begin{figure}[ht!]
    \begin{center}
\subfigure[Minimum $M_{\mathrm{KK}}$ when $U=0$]{%
            \label{fig:SParamUZero}
            \includegraphics[width=0.47\textwidth]{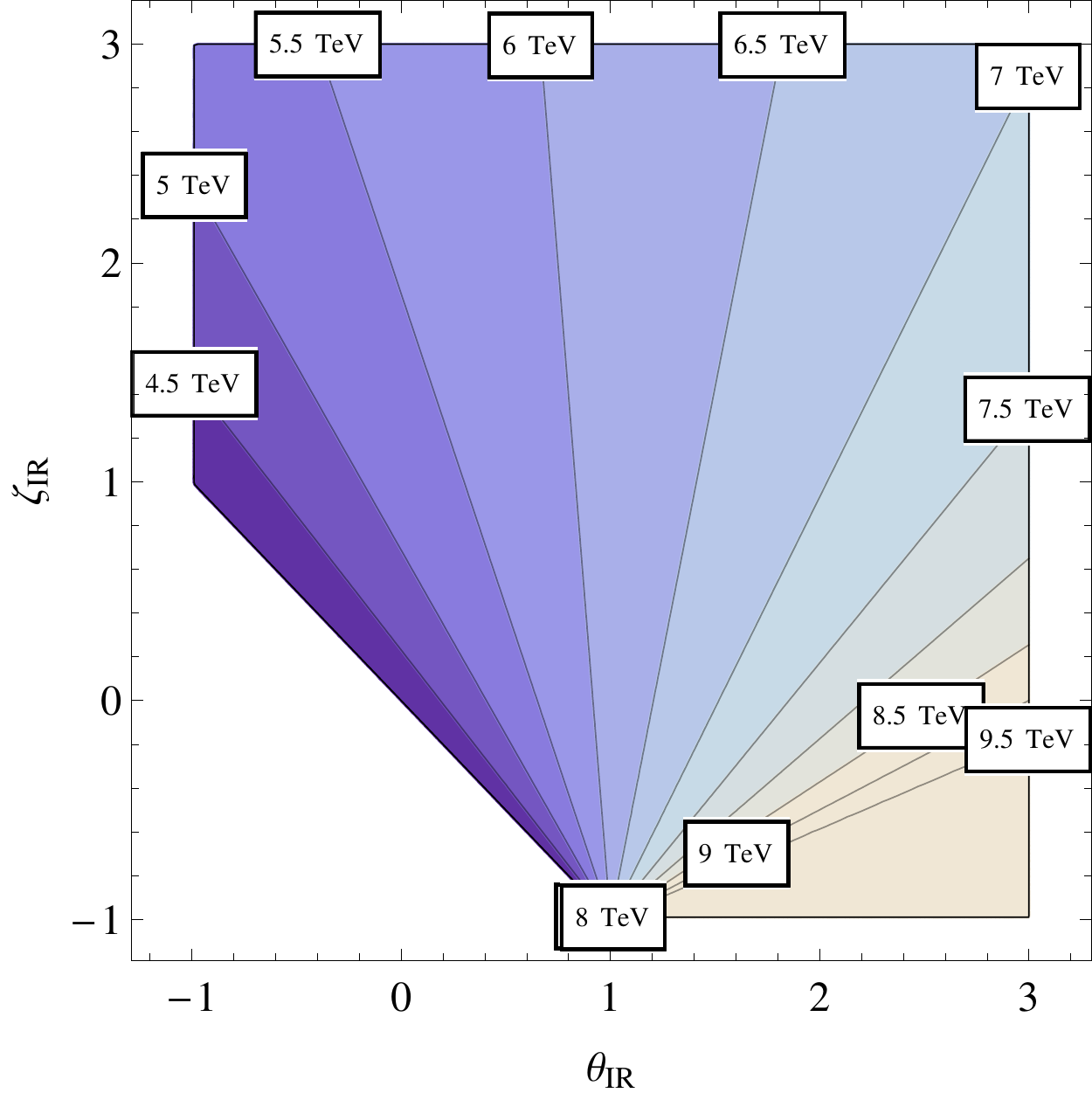}
       }
\subfigure[Minimum $M_{\mathrm{KK}}$ when $U\neq 0$]{%
            \label{fig:SParamUnonZero}
            \includegraphics[width=0.47\textwidth]{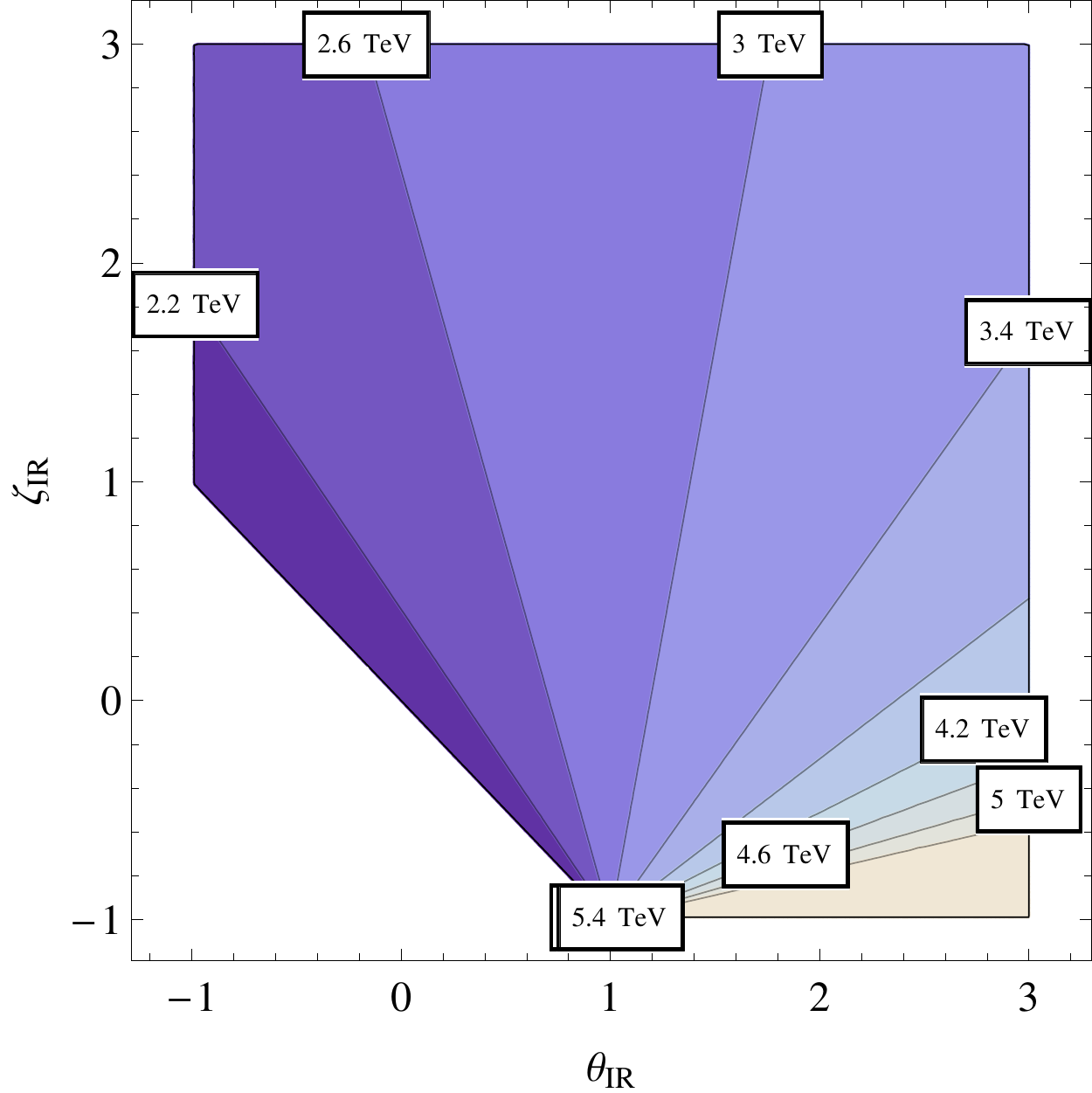}
                    }
   \end{center}
    \caption{The lower bound on $M_{\mathrm{KK}}$ from the oblique contribution to the electroweak precision tests, with $T=0$ and $\Omega=10^{15}$. The regions with $\frac{\theta_{\mathrm{IR}}+\zeta_{\mathrm{IR}}}{1+\zeta_{\mathrm{IR}}}\leqslant 0$ have been excluded due to the existence of tachyons and the constraints blow up as $\zeta_{\mathrm{IR}}\to -1$.} \label{fig:Sparameter}
\end{figure}

While the custodial symmetry does help to alleviate some of the tensions with EWPO's, it is not completely successful, since the latest fits indicate that the $S,\; T$ and $U$ parameters are quite correlated. For example, if $U=0$ the Gfitter group find that \cite{Baak:2012kk};
\begin{equation}
\label{eqn:UzeroSTparam}
S\bigg |_{U=0}=0.05\pm0.09,\hspace{1.1cm}T\bigg |_{U=0}=0.08\pm0.07\hspace{0.5cm}\mbox{ and }\hspace{0.5cm}\rho_{\mathrm{corr.}}=0.91;
\end{equation}  
indicating, at 95\% confidence level, that if $T=0$ then $-0.11\leqslant S\leqslant 0.02$. The constraints on $M_{\mathrm{KK}}$ have been plotted in figure \ref{fig:SParamUZero} and would imply (with $\Omega=10^{15}$) that $s_h^2\lesssim 0.015$ and that the mass of the first vector resonance, $m_\rho\gtrsim 10$ TeV. 

It is important to be aware of the limitations of such an analysis, particularly related to the fact that we have not included the non-oblique, $W$ and $Z$ vertex corrections and nor have we included radiative corrections which can give a sizeable contribution to the $T$ parameter \cite{Anastasiou:2009rv}. The size of such vertex corrections are sensitive to the level of compositeness of the light fermions. When the fermions are localised towards the UV brane (i.e. largely elementary), such a correction is suppressed by a factor of $1/\log (\Omega)$ and hence are numerically negligible. However, increasing the level of fermion compositeness would increase the magnitude of this correction, but would also increase the tension with flavour physics \cite{Csaki:2008zd}, as well as requiring a more detailed analysis than what is possible with just the Peskin-Takeuchi parameters. Such a study should be conducted in conjunction with a study involving all three generations and so, for reasons discussed in section \ref{sec:numerics}, lie beyond the scope of this paper (see \cite{Anastasiou:2009rv, Barbieri:2007bh, Lodone:2008yy, Gillioz:2008hs, Grojean:2013qca} for further discussion). Nonetheless, the constraints do reduce significantly if one finds any enhancement with either the $T$ or $U$ parameter. For example, if one allows for a non-zero $U$ parameter, such that \cite{Baak:2012kk};      
\begin{equation}
\label{eqn:UzeroSTparam}
S=0.03\pm0.10,\hspace{1.1cm}T=0.05\pm0.12\hspace{0.5cm}\mbox{ and }\hspace{0.5cm}\rho_{\mathrm{corr.}}=0.89;
\end{equation} 
then $T=0$ implies $-0.10\leqslant S\leqslant 0.08$ (again at 95\% confidence level), which results in the constraints plotted in figure \ref{fig:SParamUnonZero} and forces $s_h^2\lesssim 0.07$.

\subsection{The Fermion Content}
\label{sect:FermCont}
We now turn our attention to the fermion content of the model. The fermions must be embedded in representations of $\mathrm{SO}(5)$.  However, in order to avoid large corrections to the $Z\to b_L\bar{b}_L$ vertex, it is preferable to embed the SM quark  $\mathrm{SU}(2)_L$ doublets in the same representation as the singlets \cite{Agashe:2006at}. The smallest irreducible representations that permit this are the fundamental $\mathbf{5}$, the anti-symmetric $\mathbf{10}$ and the symmetric $\mathbf{14}$. Nonetheless, just considering these representations still gives rise to considerable model building possibilities, see for example \cite{Contino:2006qr, Panico:2012uw, Pappadopulo:2013vca, Carena:2014ria}. Here we wish to scrutinise the 5D realisation of the simplest model and hence we will consider the MCHM${}_5$ outlined in \cite{Contino:2006qr}.    

In the MCHM${}_5$, each SM generation is embedded in four 5D fields; $\xi_{Q_1}$ and $\xi_u$ that transform as a $\mathbf{5}_{2/3}$ of $\mathrm{SO}(5)\times\mathrm{U}(1)_X$, in addition to $\xi_{Q_2}$ and $\xi_d$ that transform as a $\mathbf{5}_{1/3}$. Upon breaking of the $\mathrm{SO}(5)$, the $\mathbf{5}$'s decompose to $\mathbf{5}=\mathbf{4}\oplus\mathbf{1}\cong(\mathbf{2},\mathbf{2})\oplus(\mathbf{1},\mathbf{1})$ under $\mathrm{SO}(4)\cong\mathrm{SU}(2)_L\times \mathrm{SU}(2)_R$. In particular, the two fields, for which we shall take the left handed component to be the source field i.e. the `$\psi$' fields (see appendix \ref{App:Fermions}), decompose as
\begin{equation}
\label{eqn:Xiq1}
\xi_{q_1}= \left[\begin{array}{cc}(\mathbf{2},\mathbf{2})_L^{q_1}=\left(\begin{array}{c}q_{1L}^\prime(-+) \\q_{1L}(++)\end{array}\right) & (\mathbf{2},\mathbf{2})_R^{q_1}=\left(\begin{array}{c}q_{1R}^\prime(+-) \\q_{1R}(--)\end{array}\right) \\(\mathbf{1},\mathbf{1})_L^{q_1}(--) & (\mathbf{1},\mathbf{1})_R^{q_1}(++)\end{array}\right] 
\end{equation}
and
\begin{equation}
\label{eqn:Xiq2}
\xi_{q_2}= \left[\begin{array}{cc}(\mathbf{2},\mathbf{2})_L^{q_2}=\left(\begin{array}{c}q_{2L}(++) \\q_{2L}^\prime(-+)\end{array}\right) & (\mathbf{2},\mathbf{2})_R^{q_2}=\left(\begin{array}{c}q_{2R}(--) \\q_{2R}^{\prime}(+-)\end{array}\right) \\(\mathbf{1},\mathbf{1})_L^{q_2}(--) & (\mathbf{1},\mathbf{1})_R^{q_2}(++)\end{array}\right]. 
\end{equation}  
where the $(\pm, \pm)$ refers to the boundary conditions on the UV and IR brane. Likewise the two `$\chi $' fields, with right handed source fields, decompose as
\begin{equation}
\label{ }
\xi_u=\left[\begin{array}{cc}(\mathbf{2},\mathbf{2})_L^u(+-) & (\mathbf{2},\mathbf{2})_R^u(-+) \\(\mathbf{1},\mathbf{1})_L^u(-+) & (\mathbf{1},\mathbf{1})_R^u(+-)\end{array}\right]\hspace{0.5cm}\mbox{ and }\hspace{0.5cm} \xi_d=\left[\begin{array}{cc}(\mathbf{2},\mathbf{2})_L^d(+-) & (\mathbf{2},\mathbf{2})_R^d(-+) \\(\mathbf{1},\mathbf{1})_L^d(-+) & (\mathbf{1},\mathbf{1})_R^d(+-)\end{array}\right].
\end{equation}      
Of central importance to the Higgs gaining a mass and a non-zero VEV, is the existence of brane localised operators that violate the $\mathrm{SO}(5)$ symmetry, analogous to the Yukawa couplings in chiral symmetry breaking. As discussed in the appendix, with the above allocation of boundary conditions, the most general set of IR operators (up to mass dimension four), that are invariant under $\mathrm{SO}(4)\times \mathrm{U}(1)_X$ and permit a low energy chiral theory, are
\begin{align}
\label{}
   \mathcal{L}_{\mathrm{IR}} =\sum_{\substack{q=q_1,q_2\;a=u,d}}&\;\theta_L^{q}R^{\prime}\overline{(\mathbf{2},\mathbf{2})_L^{q}}i\slashed{\partial} (\mathbf{2},\mathbf{2})_L^{q}+\theta_R^{q}R^{\prime}\overline{(\mathbf{1},\mathbf{1})_R^{q}}i\slashed{\partial} (\mathbf{1},\mathbf{1})^q_R+\theta_R^{a}R^{\prime}\overline{(\mathbf{2},\mathbf{2})_R^{a}}i\slashed{\partial} (\mathbf{2},\mathbf{2})_R^{a} \nonumber\\ 
   &+\theta_L^{a}R^{\prime}\overline{(\mathbf{1},\mathbf{1})_L^{a}}i\slashed{\partial}(\mathbf{1},\mathbf{1})^a_L+m_u\overline{(\mathbf{2},\mathbf{2})_L^{q_1}}(\mathbf{2},\mathbf{2})_R^{u}+M_u\overline{(\mathbf{1},\mathbf{1})_R^{q_1}}(\mathbf{1},\mathbf{1})_L^{u}\nonumber\\
   &+m_d\overline{(\mathbf{2},\mathbf{2})_L^{q_2}}(\mathbf{2},\mathbf{2})_R^{d}+M_d\overline{(\mathbf{1},\mathbf{1})_R^{q_2}}(\mathbf{1},\mathbf{1})_L^{d}+h.c. \label{MCHM5IRLag}  
\end{align}  
In theory, one should also include an analogous set of operators on the UV brane. Although, from our experience with the gauge sector, one finds that when $\Omega$ is large, UV localised operators have a relatively small effect on EW scale physics. Hence we will simplify the scenario and just consider IR localised operators. Previous studies, for simplicity, have often just included the brane localised mass terms although in doing so one is assuming that the unknown physics, responsible for the Higgs mass and EWSB, carries no momentum dependence.     

Following \cite{Contino:2006qr}, one finds, after projecting out the spurious fields with Dirichlet UV BC's,  that the holographic effective Lagrangian is given by, see appendix \ref{app:Fermion},
\begin{align}
\label{FermLagrangian}
\mathcal{L}_{\mathrm{hol.}}&=\bar{q}_L\left [\Pi_0^q+\frac{s_h^2}{2}\left(\Pi_1^{q1}H^cH^{c\dag}+\Pi_1^{q2}HH^{\dag} \right )\right ]\slashed{p}q_L+\sum_{r=u,d}\bar{r}_R\left [\Pi_0^r+\frac{s_h^2}{2}\Pi_1^r\right ]\slashed{p}r_R\nonumber\\
&\hspace{5.5cm}+\frac{s_hc_h}{\sqrt{2}}\left (M_1^u\bar{q}_LH^cu_R+M_1^d\bar{q}_LHd_R\right )+h.c,
\end{align} 
where
\begin{equation}
\label{ Hdef}
H=\frac{1}{h}\left (\begin{array}{c}h^1-ih^2 \\ h^3-ih^4\end{array}\right )\quad\mbox{ and }\quad H^c=\frac{1}{h}\left(\begin{array}{c}-(h^1+ih^2) \\h^3+ih^4\end{array}\right)
\end{equation}
and the $\Pi_{0,1}$ form factors are specified in (\ref{eqn:FermPis}). The above expression should carry flavour indices. Since this paper is primarily focused with the tuning in the Higgs potential, of which the fermion contribution is approximately proportional to the the mass of the fermion zero mode (or equivalently it's level of compositeness), from this point onwards we shall just consider the top and bottom quarks and neglect the first two generations, in addition to the leptons. 

After EWSB and canonically normalising the fields, the SM top and bottom masses are given, at zero momentum, by
\begin{eqnarray}
m_t & = & \frac{s_hc_h}{\sqrt{2}}\frac{M_1^u(0)}{\sqrt{\Pi_0^q(0)+\frac{s_h^2}{2}\Pi_1^{q1}(0)}\sqrt{\Pi_0^u(0)+\frac{s_h^2}{2}\Pi_1^{u}(0)}}, \\
m_b& = & \frac{s_hc_h}{\sqrt{2}}\frac{M_1^d(0)}{\sqrt{\Pi_0^q(0)+\frac{s_h^2}{2}\Pi_1^{q2}(0)}\sqrt{\Pi_0^d(0)+\frac{s_h^2}{2}\Pi_1^{u}(0)}} .
\end{eqnarray}
While there also exists;
\begin{itemize}
  \item  a tower of charge +2/3 fermions with masses given by: 
  \begin{displaymath}
\mathrm{zeros}\left \{p^2\left (\Pi_0^q+\frac{s_h^2}{2}\Pi_1^{q1}\right )\left (\Pi_0^u+\frac{s_h^2}{2}\Pi_1^{u}\right )-\frac{s_h^2c_h^2}{2}\left ( M_1^u\right )^2\right \}
\end{displaymath}
  \item a tower of charge -1/3 fermions with masses given by: 
  \begin{displaymath}
\mathrm{zeros}\left \{p^2\left (\Pi_0^q+\frac{s_h^2}{2}\Pi_1^{q1}\right )\left (\Pi_0^u+\frac{s_h^2}{2}\Pi_1^{u}\right )-\frac{s_h^2c_h^2}{2}\left ( M_1^u\right )^2\right \}
\end{displaymath}
  \item a tower of charge +5/3 fermions with masses given by: 
    \begin{displaymath}
\mathrm{poles}\left \{\slashed{p}\left ( \Pi_0^u+\frac{1}{2}\Pi_1^u\right )\right \}
\end{displaymath}
 \item a tower of charge -4/3 fermions with masses given by: 
    \begin{displaymath}
\mathrm{poles}\left \{\slashed{p}\left ( \Pi_0^d+\frac{1}{2}\Pi_1^d\right )\right \}
\end{displaymath}
\end{itemize}

\section{The Higgs potential}
\label{sec:higgspot}
Having outlined the MCHM, we will now move on to the central question of this paper, notably how much fine tuning is required in order to obtain the correct Higgs VEV and mass, given that no new heavy resonances have currently been observed? Central to this question is the analysis of the Higgs potential, which in the 5D holographic MCHM is generated via the Coleman-Weinberg mechanism \cite{Coleman:1973jx}. Here we compute the leading order contribution to the Higgs potential, although \cite{Barnard:2013hka} found that the next to leading order fermion contribution can be sizeable in these models. It is also worth commenting, in models of gauge-Higgs unification in more than five dimensions, the Higgs potential would naively be generated at tree level and so generating viable models models, with fermion masses and minimal tuning, is more challenging \cite{Antoniadis:2001cv, Lim:2006bx}.   

\subsection{The Gauge Contribution}
The Higgs potential will receive contributions from both the gauge fields and the fermion fields. If we start
by considering the gauge sector, described by 
\begin{displaymath}
\mathcal{L}_{\mathrm{eff.}}=-\frac{P_t^{\mu\nu}}{2g^2}\;A_\mu^{\hat{a}} \Pi_G^{\hat{a}\hat{b}}(p)A_\nu^{\hat{b}}
\end{displaymath}
with $A_\mu^{\hat{a}}=\{A_\mu, W_\mu^\pm, Z_\mu\}$, then the Higgs potential will be given by \cite{Coleman:1973jx, Serone:2009kf}
\begin{equation}
\label{eqn:CWGaugePot}
V_G(s_h^2)=\frac{3}{2}\int \frac{d^4p_E}{(2\pi)^4}\log \left [ \det\left (\Pi_G(p_E)\right )\right ].
\end{equation}
After substituting in (\ref{eqn:LagPostEWSB}) and absorbing the logarithmically divergent terms by wavefunction renormalisation, then the part of the Higgs potential of relevance to determining $s_h^2$ is found to be
\begin{eqnarray}
V_G(s_h^2) & = & \frac{3}{2}\int \frac{d^4p_E}{(2\pi)^4}\log\left[1+\left (\frac{3\Pi_1}{2\Pi_0}+\frac{s_x^2\Pi_1}{2\Pi_B}\right )s_h^2+\left (\frac{\Pi_1\left (s_x^2\Pi_0\Pi_1+\Pi_B\Pi_1\right )}{2\Pi_B\Pi_0^2}+\frac{\Pi_1^2}{4\Pi_0^2}\right )s_h^4\right ] \nonumber\\
 & \approx & \frac{3}{2}\int \frac{d^4p_E}{(2\pi)^4}\left [\left (\frac{3\Pi_1}{2\Pi_0}+\frac{s_x^2\Pi_1}{2\Pi_B}\right )s_h^2-\left (\frac{3}{8\Pi_0^2}+\frac{s_x^2}{4\Pi_0\Pi_B}+\frac{s_x^2}{8\Pi_B^2}\right )\Pi_1^2s_h^4+\mathcal{O}(s_h^6)\right ] \nonumber\\
 & = &-\alpha_Gs_h^2+\beta_Gs_h^4+\dots \label{eqn:GaugeHiggsPotential}
\end{eqnarray}
where $\Pi_0=\Pi^{(+)}(p_E)$, $\Pi_1=\Pi^{(-)}(p_E)-\Pi^{(+)}(p_E)$ and $\Pi_B=s_x^2\Pi^{(+)}(p_E)+c_x^2\Pi_X^{(+)}(p_E)$ and the form factors are given in (\ref{eqn:GaugePiNegEuclid} \& \ref{eqn:GaugePiPosEuclid}). In the third line we have adopted a standard parameterisation of the Higgs potential. Note that since the integrand cannot change sign, without giving rise to a tachyonic mode, $\alpha_G$ and $\beta_G$ are always negative.   

\subsection{The Fermion Contribution}
Likewise the fermion contribution, from the effective Lagrangian
\begin{displaymath}
\mathcal{L}=\bar{\Psi}^I\Pi_\Psi^{IJ}(p)\Psi^J,
\end{displaymath}
is given by
\begin{equation}
\label{CWPsiPot}
V_\Psi(s_h^2)=-2\int \frac{d^4p_E}{(2\pi)^4}\;\log\left [\det\left (\Pi_\Psi(p_E)\right )\right ].
\end{equation}
Which for (\ref{FermLagrangian}) leads to the relevant contribution from the top quarks being (again after absorbing the logarithmically divergent part in the wavefunction renormalisation),
\begin{eqnarray}
V_t(s_h^2) & = & -2N_C\int\frac{d^4p_E}{(2\pi)^4}\;\log\left [\left (1+\frac{s_h^2}{2}\frac{\Pi_1^{q1}}{\Pi_0^q}\right )\left (1+\frac{s_h^2}{2}\frac{\Pi_1^{u}}{\Pi_0^u}\right )+\frac{s_h^2c_h^2}{2}\frac{(M_1^u)^2}{p_E^2\Pi_0^q\Pi_0^u}\right ] \nonumber\\
 & \approx &  -2N_C\int\frac{d^4p_E}{(2\pi)^4}\;\left (\frac{\Pi_1^q}{2\Pi_0^q}+\frac{\Pi_1^u}{2\Pi_0^u}+\frac{(M_1^u)^2}{2p_E^2\Pi_0^q\Pi_0^u}\right )s_h^2\nonumber\\
 &&\hspace{0.5cm}+N_C\int\frac{d^4p_E}{(2\pi)^4}\left [\left (\frac{\Pi_1^q}{2\Pi_0^q}+\frac{\Pi_1^u}{2\Pi_0^u}+\frac{(M_1^u)^2}{2p_E^2\Pi_0^q\Pi_0^u}\right )^2+\frac{(M_1^u)^2}{p_E^2\Pi_0^q\Pi_0^u}-\frac{\Pi_1^q\Pi_1^u}{2\Pi_0^q\Pi_0^u}\right ]s_h^4+\mathcal{O}(s_h^6)\nonumber\\
 &\equiv& -\alpha_t s_h^2+\beta_t s_h^4+\dots \label{eqn:FermHiggsPotential}
\end{eqnarray}
where $N_C=3$ is the number of QCD colours. An analogous expression clearly exists for the bottom quark contribution, $\alpha_b$ and $\beta_b$. Note that now $\alpha_{t,b}$ and $\beta_{t,b}$ are now positive in viable models. 

All of the integrands, in $\alpha_{G,t,b}$ and $\beta_{G,t,b}$, go to zero for large momenta and satisfy the sum rules used in \cite{Marzocca:2012zn, Pomarol:2012qf}. In fact for $\alpha_{G}$ and $\beta_{G}$ this can be shown analytically using the Wronskian relation, $\mathbf{I}_0(x)\mathbf{K}_1(x)+\mathbf{K}_0(x)\mathbf{I}_1(x)=1/x$. Although, as previously noted in \cite{Marzocca:2012zn}, there is spurious IR divergence in $\beta_{G,t,b}$ arising from the expansion of the $\log$. This is removed with an IR cut-off at $\Lambda_{\mathrm{IR}}=M_{\mathrm{KK}}/100$, but the results are insensitive to this cut-off. The integrands decrease exponentially with respect to $p_E$ and typically become small for $p_E\gtrsim 4-7\; M_{\mathrm{KK}}$, i.e. lower than the UV cut-off $\sim 10\;M_{\mathrm{KK}}$ \cite{Csaki:2008zd}. Hence one can have reasonable confidence in the numerical evaluation of the integrals.

\subsection{Analysis of the Higgs Potential}

Having found the Higgs potential, expanded around $s_h^2$, 
\begin{equation}
\label{HiggsPotFull}
V(s_h^2)\approx -\alpha s_h^2+\beta s_h^4+\mathcal{O}(s_h^6),
\end{equation}  
where $\alpha=\alpha_G+\alpha_b+\alpha_t$ and $\beta=\beta_G+\beta_b+\beta_t$, it is straightforward to find the expectation value of $s_h^2$, that we require to be (\ref{eqn:sh2 }),
\begin{equation}
\label{sh2fromPotential}
s_h^2\approx\frac{\alpha}{2\beta}.
\end{equation}
 Likewise the Higgs mass can also be found from the potential,
 \begin{equation}
\label{eqn:MH2}
m_H^2=\frac{V^{\prime\prime}(H)}{f_\pi^2}\bigg |_{H=\langle H\rangle}\approx \frac{2\alpha(2\beta-\alpha)}{\beta f_\pi^2}=\frac{8\beta s_h^2c_h^2}{f_\pi^2}.
\end{equation} 
In order to achieve viable EWSB one requires that
\begin{equation}
\label{ViableEWSBCons}
2\beta>\alpha>0,
\end{equation}
which requires that the fermion contribution, to the potential, dominates over the gauge contribution. It follows, from the above relations that we require that
\begin{displaymath}
\alpha\approx\frac{f_\pi^2m_H^2}{2c_w^2}\hspace{0.7cm}\mbox{ and }\hspace{0.7cm}\beta\approx\frac{f_\pi^4m_H^2
}{8v^2c_w^2}.
\end{displaymath}
When the fermions are embedded in fundamental $\mathbf{5}$'s (or anti-symmetric $\mathbf{10}$'s), one typically finds that $\alpha_{t,b} > \beta_{t,b}$. This can be most elegantly understood by promoting the effective Yukawa couplings to spurions and noting that the $\alpha_{t,b}$ arises from one spurion insertion, whereas $\beta_{t,b}$ requires two insertions \cite{Matsedonskyi:2012ym, Panico:2012uw}. This is also the reason why the integrand in $\beta_{t,b}$ is related to the square of the integrand of $\alpha_{t,b}$ in (\ref{eqn:FermHiggsPotential}). While this unhelpful feature can be avoided by considering larger fermion representations, such as $\mathbf{14}$'s \cite{Panico:2012uw, Pappadopulo:2013vca}, for the MCHM${}_5$ it gives rise to a significant source of fine tuning. In particular, in order to realise (\ref{ViableEWSBCons}), it is not only necessary for the fermion contribution to dominate over the gauge contribution, but also that $\alpha_G$ must cancel with $\alpha_{t}+\alpha_{b}$, such that $\alpha\ll \alpha_{t}$.      

\subsection{The Study of a Benchmark Point}

\begin{figure}[ht!]
    \begin{center}
  \subfigure[$m_t$ upon variation of $\theta_R^u$ and $\theta_L^{q_1}$]{%
            \label{fig:mt_Benchmark}
            \includegraphics[width=0.375\textwidth]{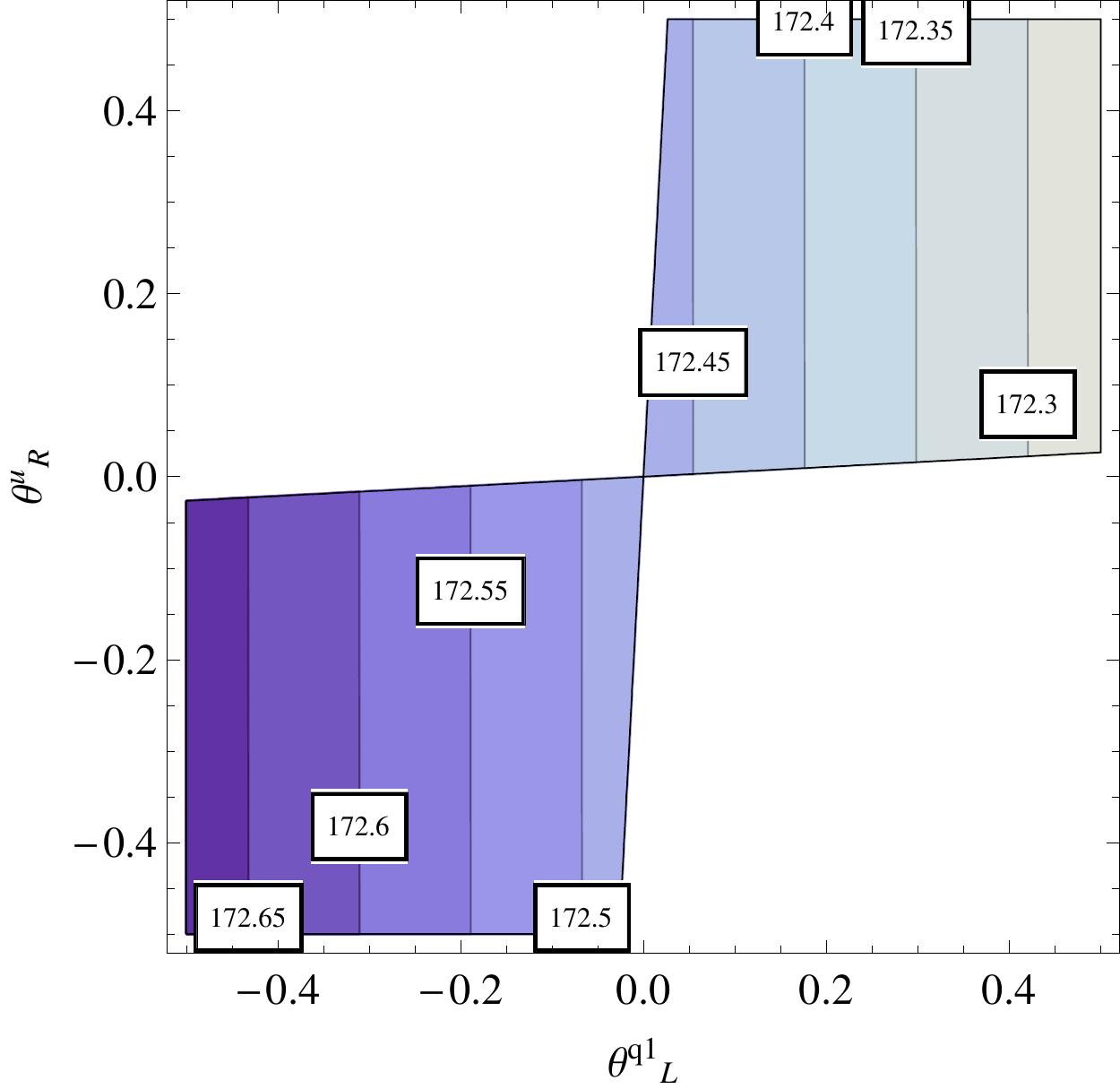}
       }\\  
\subfigure[$\frac{\alpha_t+\alpha_b}{\alpha_0}$ upon variation of $\theta_R^u$ and $\theta_L^{q_1}$]{%
            \label{fig:alphaT_Benchmark}
            \includegraphics[width=0.375\textwidth]{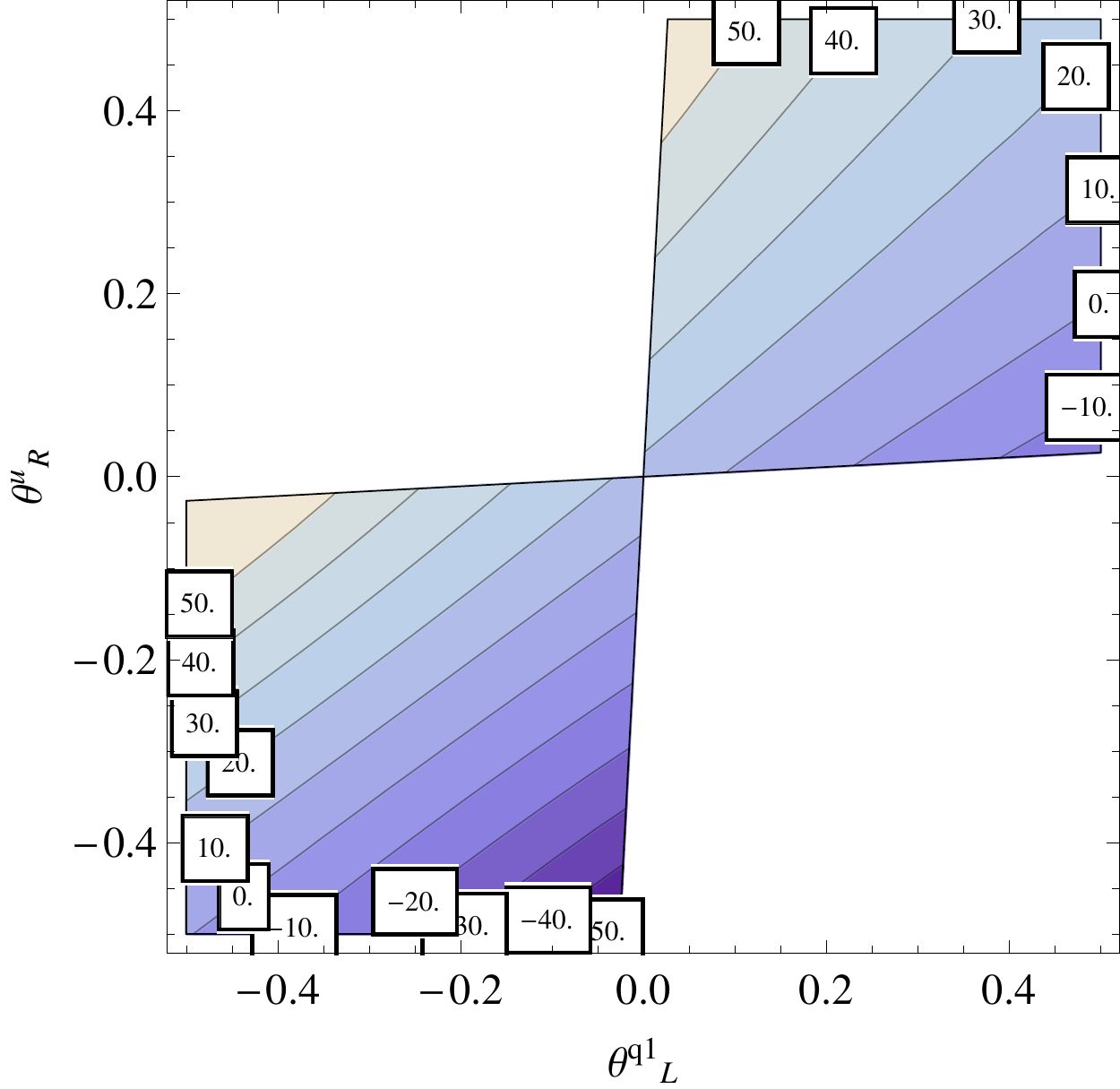}
       }
\subfigure[$\frac{\beta_t+\beta_b}{\alpha_0}$ upon variation of $\theta_R^u$ and $\theta_L^{q_1}$]{%
            \label{fig:betat_Benchmark}
            \includegraphics[width=0.375\textwidth]{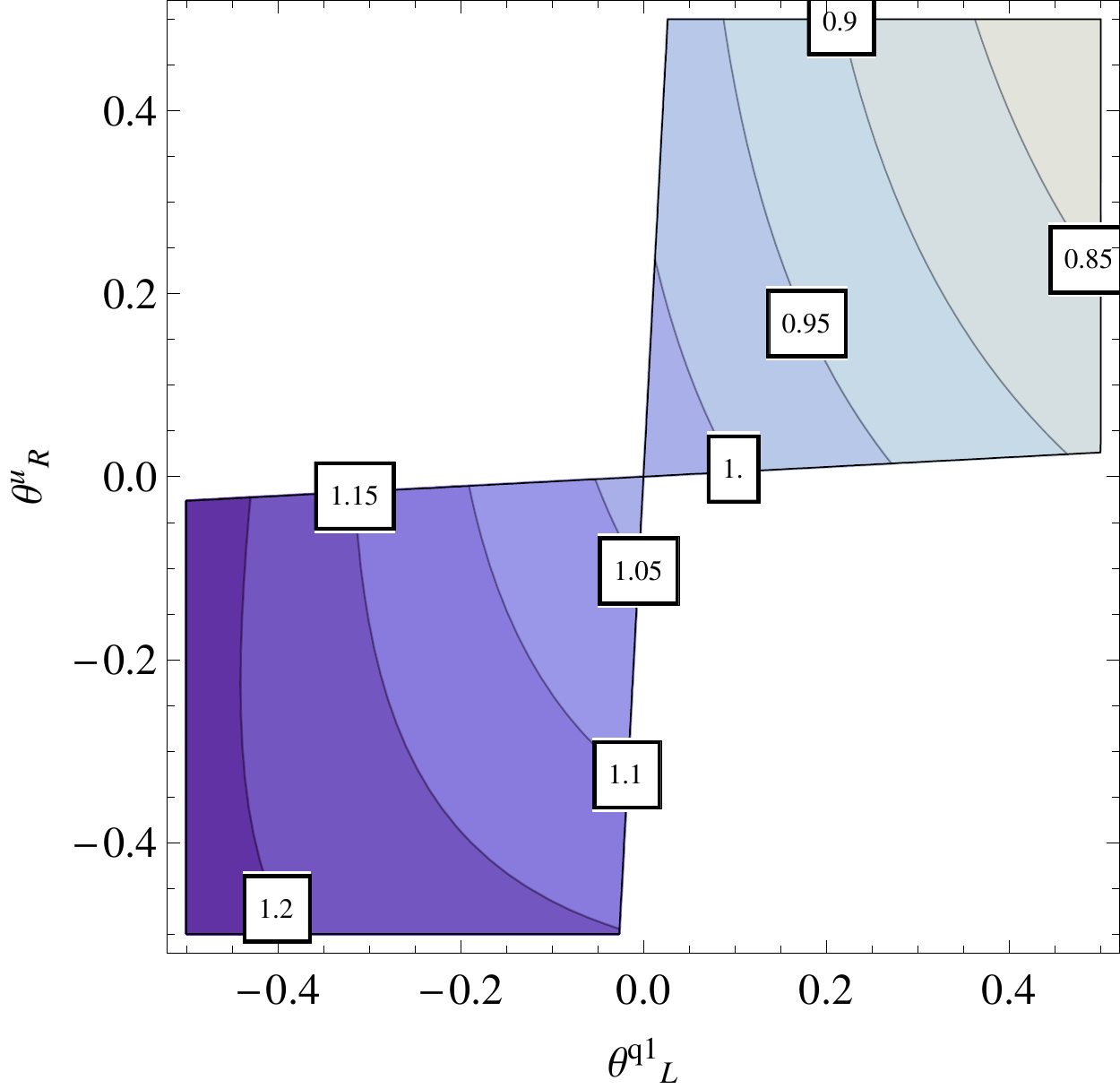}
                    }\\              
 \subfigure[$\frac{\alpha_G}{\alpha_0}$ upon variation of $\zeta_{\mathrm{IR}}$ and $\theta_{\mathrm{IR}}$]{%
            \label{fig:AlphaG_Benchmark}
            \includegraphics[width=0.375\textwidth]{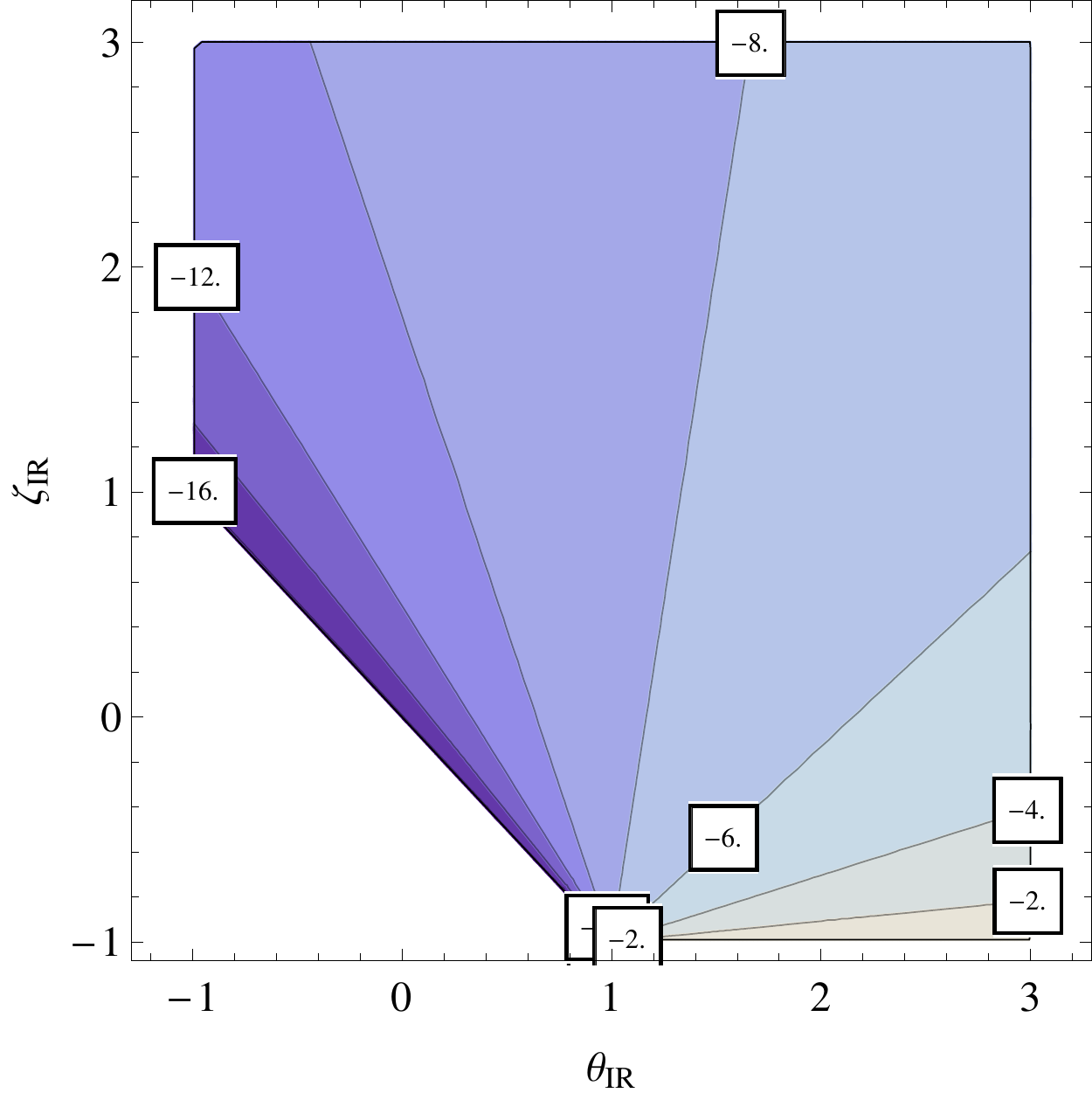}
       }
\subfigure[$\frac{\beta_G}{\beta_0}$ upon variation of $\zeta_{\mathrm{IR}}$ and $\theta_{\mathrm{IR}}$]{%
            \label{fig:betaG_Benchmark}
            \includegraphics[width=0.375\textwidth]{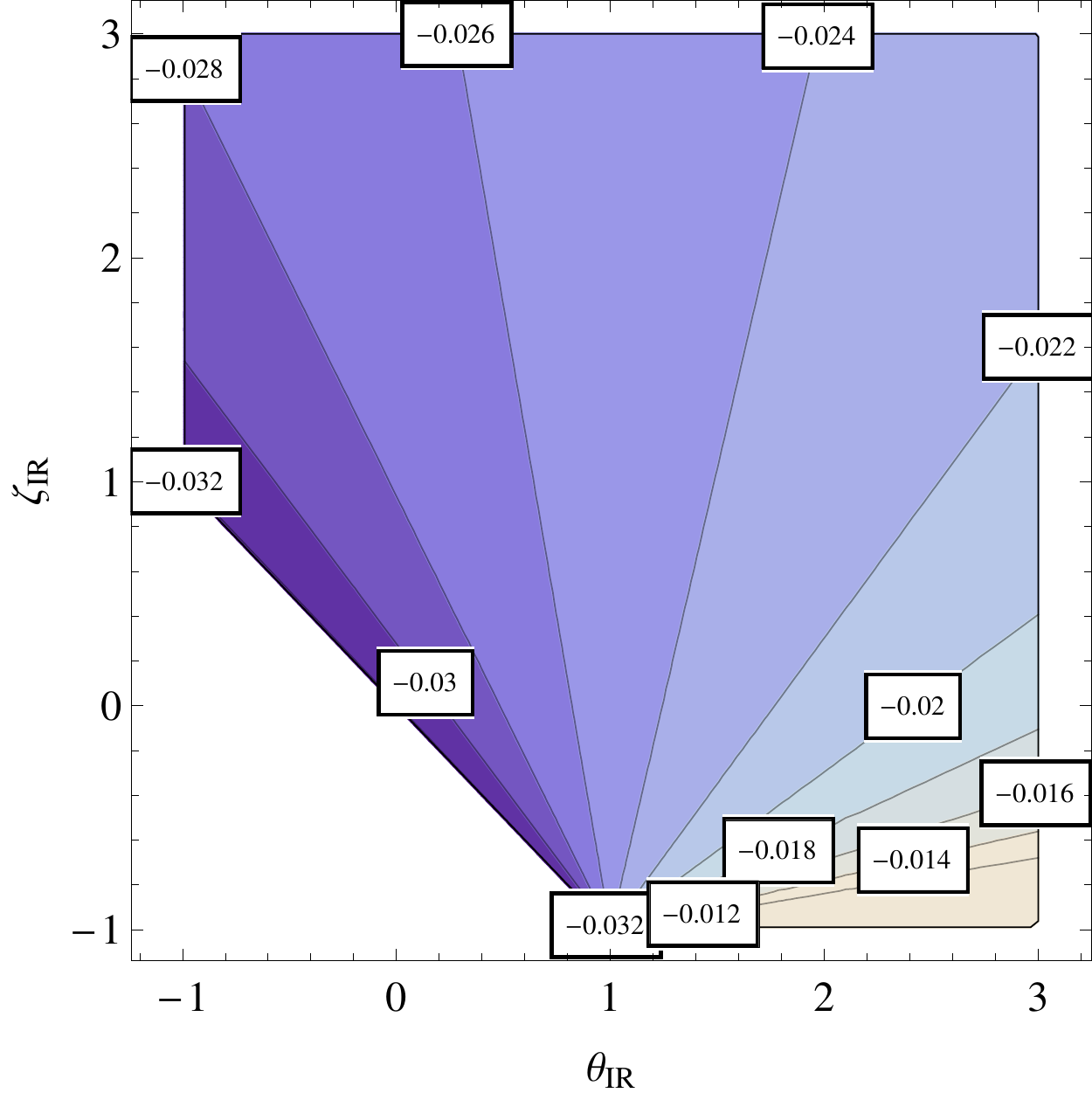}
          }
   \end{center}
    \caption{Variation of the Higgs potential coefficients relative to $\alpha_0$ and $\beta_0$, the value of the coefficients with the input parameters (\protect\ref{BenchmarkInputParameters}). The white region has been excluded by the existence of tachyon modes, see (\protect\ref{eqn:TachonCond} \& \protect\ref{FermionTacyonConstraint}).}\label{fig:BenchMark}
\end{figure}

In order to emphasise this fine tuning, it is productive to study a specific viable point in parameter space, found in  section \ref{sec:numerics}.  The point we arbitrarily choose has the input parameters,
\begin{equation}
\label{BenchmarkInputParameters}
\arraycolsep=8pt\def\arraystretch{1}
\begin{array}{lllll}
  M_{\mathrm{KK}}=2.44\; \mathrm{TeV}, & c_{q_1}=-0.021,& c_{q_2}=0.071,& c_{u}=0.21,& c_{d}=0.48, \\
  m_u=-2.06, & m_d=2.56, & M_u=0.013, & M_d=0.086, & \\
  \theta_{\mathrm{IR}}=\zeta_{\mathrm{IR}}=0,& \theta^{(X)}_{\mathrm{IR}}=\zeta^{(X)}_{\mathrm{IR}}=0,&\theta_L^{q_1,\; q_2,\;u,\;d}=0,&\theta_R^{q_1,\; q_2,\;u,\;d}=0,&
\end{array}
\end{equation}
which gives rise to
\begin{displaymath}
\arraycolsep=8pt\def\arraystretch{1}
\begin{array}{lllll}
      m_t=172.5\; \mathrm{GeV}, & m_b=4.21\; \mathrm{GeV}, &m_h=125.8\; \mathrm{GeV}, &s_h^2=0.037,  & \Delta_{\mathrm{B.G}}=43   
\end{array}
\end{displaymath}
as well as the lightest fermion resonances with charge $Q$, having masses $m_Q$
\begin{displaymath}
\arraycolsep=8pt\def\arraystretch{1}
\begin{array}{llll}
      m_{+\frac{2}{3}}=3.70\; \mathrm{TeV}, & m_{-\frac{1}{3}}=3.71\; \mathrm{TeV}, &m_{+\frac{5}{3}}=0.82\; \mathrm{TeV},&m_{-\frac{4}{3}}=5.98\; \mathrm{TeV}.  
\end{array}
\end{displaymath} 
Here we have, at random taken a point in parameter space, without brane localised kinetic terms, that satisfies all of the experimental constraints (assuming $U\neq 0$). While there is nothing remarkable about this point, it does allow us to emphasise a few important effects, by `turning on' the IR localised kinetic terms and taking a slice through the parameter space, see figure \ref{fig:BenchMark}:
\begin{itemize}
  \item Firstly, that the coefficients of $\alpha$ and $\beta$ can be enhanced or suppressed by including brane localised kinetic terms and that this is not necessarily correlated with shifts in $m_t$ and $m_b$, see figures \ref{fig:mt_Benchmark}, \ref{fig:alphaT_Benchmark} and \ref{fig:betat_Benchmark}.   
  \item However, comparing figure \ref{fig:Sparameter} with \ref{fig:AlphaG_Benchmark} and \ref{fig:betaG_Benchmark}, shifts in $\alpha_G$ and $\beta_G$ are precisely correlated with shifts in the $S$ parameter. One finds that, with the brane localised kinetic terms set to zero, the contribution to the $S$ parameter is at its minimal value and the coefficients $\alpha_G$ and $\beta_G$ are at their maximal values. 
  \item Due to the reasons discussed in the previous section, for viable points satisfying (\ref{ViableEWSBCons}), the relative size of the fermion and gauge contributions differs significantly for $\alpha$ and for $\beta$. For $\beta$ the fermion contribution dominates over the gauge contribution. This is partly due to the factor of $N_C$ enhancement in (\ref{eqn:FermHiggsPotential}) compared to (\ref{eqn:GaugeHiggsPotential}).  
  \item Whereas $\alpha_G$ must necessarily be of a similar size to $\alpha_t+\alpha_b$ in order for the two coefficients to cancel and realise (\ref{ViableEWSBCons}).     
\end{itemize}  
This final point represents the dominant source of fine tuning in the model. As has already been mentioned, this fine tuning can be reduced by considering alternative fermion representations \cite{Panico:2012uw, Pappadopulo:2013vca}. Although it has also been found that the overall tuning can increase, due to such models preferring a heavier Higgs \cite{Carena:2014ria}. Either way, what is initially required is a quantitative understanding of how much fine tuning is required in the MCHM, with a minimal fermion content. So, we shall now attempt a detailed numerical analysis of the MCHM${}_5$.

\section{Numerical Analysis}
\label{sec:numerics}

What should have been demonstrated, in the previous section, is that the MCHM${}_5$ has a large and under-constrained parameter space. Since the 5D theory is an effective theory, a priori there is no reason to set the brane localised kinetic terms to zero. So the relevant question is firstly, is it possible to find a viable point in parameter space and secondly, is such a point `natural'? Here a viable point is a point that gives rise to EWSB with the correct masses for all the observed particles. Given the size of the parameter space, one would expect the answer to the first question to be yes, but the second question is more challenging since defining what makes a theory natural is difficult. One of the most common criteria, used for measuring the naturalness of a theory, is to look at the level of fine tuning that exists. For a model, with the observables $\mathcal{O}_i$,  determined by the input parameters $X_j$, the level of fine tuning is conventionally defined to be \cite{Barbieri:1987fn}
\begin{equation}
\label{FTparam}
\Delta_{\mathrm{B.G}}=\max_{i,j}\left |\frac{X_j}{\mathcal{O}_i}\frac{\partial \mathcal{O}_i}{\partial X_j}\right |\approx \max_{i,j}\left |\frac{X_j}{\mathcal{O}_i}\frac{\Delta \mathcal{O}_i}{\Delta X_j}\right |,
\end{equation}      
although such a definition is not perfect. Firstly, some observables would result in a large fine tuning no matter what its value was. A famous example is the sensitivity of the proton mass to the QCD cut-off \cite{Anderson:1994dz}, while other such examples in the MCHM would include the sensitivity of the SM fermion masses to the bulk mass parameters ($c_i$). It is important to distinguish these cases from observables, such as the Higgs mass, that arise from the additive cancellation between two parameters. Fortunately, in practice here the fine tuning of the top and bottom mass is typically smaller than that of the Higgs mass and VEV and so we can safely ignore such subtleties.        

Another issue is that fine tuning does not give a measure of the rarity of the point in parameter space. Should, for example, a theory with a single small `sweet-spot' with low fine-tuning, with the rest of the parameter space having high levels fine tuning, be considered natural? With such questions in mind, here we shall attempt to be relatively impartial in our scan over the parameter space. 

\subsection{The Scan Over Parameter Space} 
\label{sec:ParamScan}    
 In order to study the level of fine tuning, we must first find viable points. Due to the computationally slow numerical integrals in (\ref{eqn:FermHiggsPotential}), this is a non-trivial optimisation problem. However since it is found that the contributions, to the Higgs potential, scale approximately with the mass of the fermion zero modes, we simplify the model by considering just the top and bottom quark. This reduces the size of the parameter space and the number of integrals to compute. Another issue arises since $s_h^2$ appears as both an input and an output parameter. Here, for the computation of the Higgs potential and quark masses, we use the value of $s_h^2$ required to give the correct W and Z mass (\ref{eqn:sh2 }), before fitting the true value ($\alpha / 2\beta$) to this required value. In addition to $s_h^2$, we also fit to\footnote{Here we encounter a problem with the analysis, since it is unlikely that the quark and Higgs masses would run as they do in the SM and in addition we are dealing with observables measured at multiple energy scales. It can be checked that the results do not change significantly if the input quark masses are varied by a few per cent and so here we fit, at low momentum, to the top pole mass and $m_b(\overline{\mathrm{MS}})$.}   
 \begin{equation}
\label{ }
m_t=173\pm0.88\;\mbox{GeV},\hspace{0.5cm}m_b=4.18\pm0.03\;\mbox{GeV},\hspace{0.5cm}m_h=125.9\pm 0.4\;\mbox{GeV}. 
\end{equation}
 The fit is made by minimising some function\footnote{If the model gives the value $X_i$ for an observable, measured to be $\mathcal{O}_i\pm\Delta\mathcal{O}_i$, we minimise $\sum_i\prod_{j\neq i}\frac{\mathcal{O}_j^2}{X_j^2}\frac{(\mathcal{O}_i-X_i)^2}{\Delta\mathcal{O}_i^2}$. This avoids the algorithm getting stuck in a local minima with $X_i\to 0$.} using a simulated annealing algorithm starting from a randomly chosen initial point, with the input parameters
 \begin{equation}
\label{InitialValuesInputParameters}
\arraycolsep=8pt\def\arraystretch{1}
\begin{array}{llll}
  M_{\mathrm{KK}}\in[0.5,\;4]\; \mathrm{TeV}, & c_{q_1,q_2}\in[0.6,\;0],& c_{u,d}\in[-0.6,\;0],& m_{u,d}\in[-3,\;3],\\
   M_{u,d}\in[-3,\;3],&\theta_{L,R}^{q_1,\; q_2,\;u,\;d}\in[-\theta_\psi,\;\theta_\psi], & \theta^{(X)}_{\mathrm{IR}}=\zeta^{(X)}_{\mathrm{IR}}=0.&
 \end{array}
\end{equation}
 We will vary the parameters $\theta_\psi$, $\theta_{\mathrm{IR}}$ and $\zeta_{\mathrm{IR}}$ in order to consider different regions of the parameter space. The fine tuning parameter is computed using (\ref{FTparam}) and perturbing all of the input parameters by a random amount of up to 2.5\%, 5\% and 10\%, before extrapolating to 0\%. This is found to give a relatively stable number.  

In order to investigate the effect of such brane localised kinetic terms we will consider 5 scenarios:
\begin{itemize}
  \item No kinetic terms, i.e. $\theta_\psi=\theta_{\mathrm{IR}}=\zeta_{\mathrm{IR}}=0$.
  \item A partially suppressed gauge contribution with $\theta_{\mathrm{IR}}=1$, $\zeta_{\mathrm{IR}}=1$ and $\theta_\psi=0$. 
  \item  A suppressed gauge contribution with $\theta_{\mathrm{IR}}=3$, $\zeta_{\mathrm{IR}}=0$ and $\theta_\psi=0$. 
  \item Small anarchic fermion kinetic terms,  $\theta_\psi=0.5$, with $\theta_{\mathrm{IR}}=\zeta_{\mathrm{IR}}=0$.
  \item Large anarchic fermion kinetic terms,  $\theta_\psi=1.5$, with $\theta_{\mathrm{IR}}=\zeta_{\mathrm{IR}}=0$.
\end{itemize}
For each scenario, we run the optimisation routine 1500 times and select points which have converged to an extent such that $s_h^2$ lies within 30\% of the required value and $m_h$, $m_t$ and $m_b$ are correct to within 10\%. Typically all points that satisfy the former constraint also satisfy the latter constraint. It is also worth commenting that, for the viable points, there is a high correlation between $c_{q_1}$ and $c_{q_2}$, i.e. $c_{q_1}\approx c_{q_2}$. This correlation does not exist between the other bulk mass parameters.

 \subsection{The Fine Tuning Parameter}
 
 \begin{figure}
\begin{center}

 \subfigure[Fine tuning for $\theta_\psi=\theta_{\mathrm{IR}}=\zeta_{\mathrm{IR}}=0$ (green), \newline $\theta_{\mathrm{IR}}=\zeta_{\mathrm{IR}}=1$, $\theta_\psi=0$ (blue) and\newline $\theta_{\mathrm{IR}}=3$, $\zeta_{\mathrm{IR}}=\theta_\psi=0$ (red).]{%
            \label{fig:FTvMKKgauge}
            \includegraphics[width=0.45\textwidth]{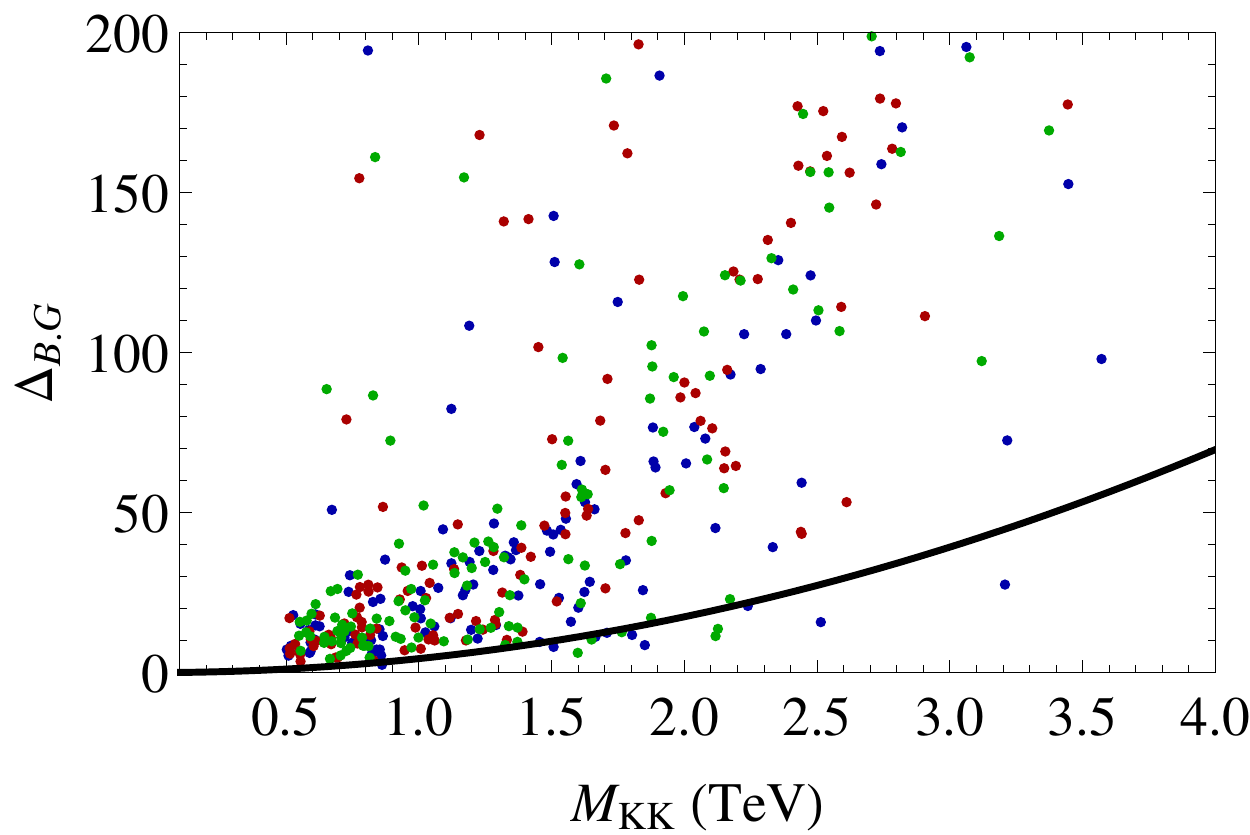}
       }
\subfigure[Fine tuning for $\theta_\psi=0.5$, $\theta_{\mathrm{IR}}=\zeta_{\mathrm{IR}}=0$ (blue)\newline and $\theta_\psi=1.5$, $\theta_{\mathrm{IR}}=\zeta_{\mathrm{IR}}=0$ (red). ]{%
            \label{fig:FTvMKKgauge}
            \includegraphics[width=0.45\textwidth]{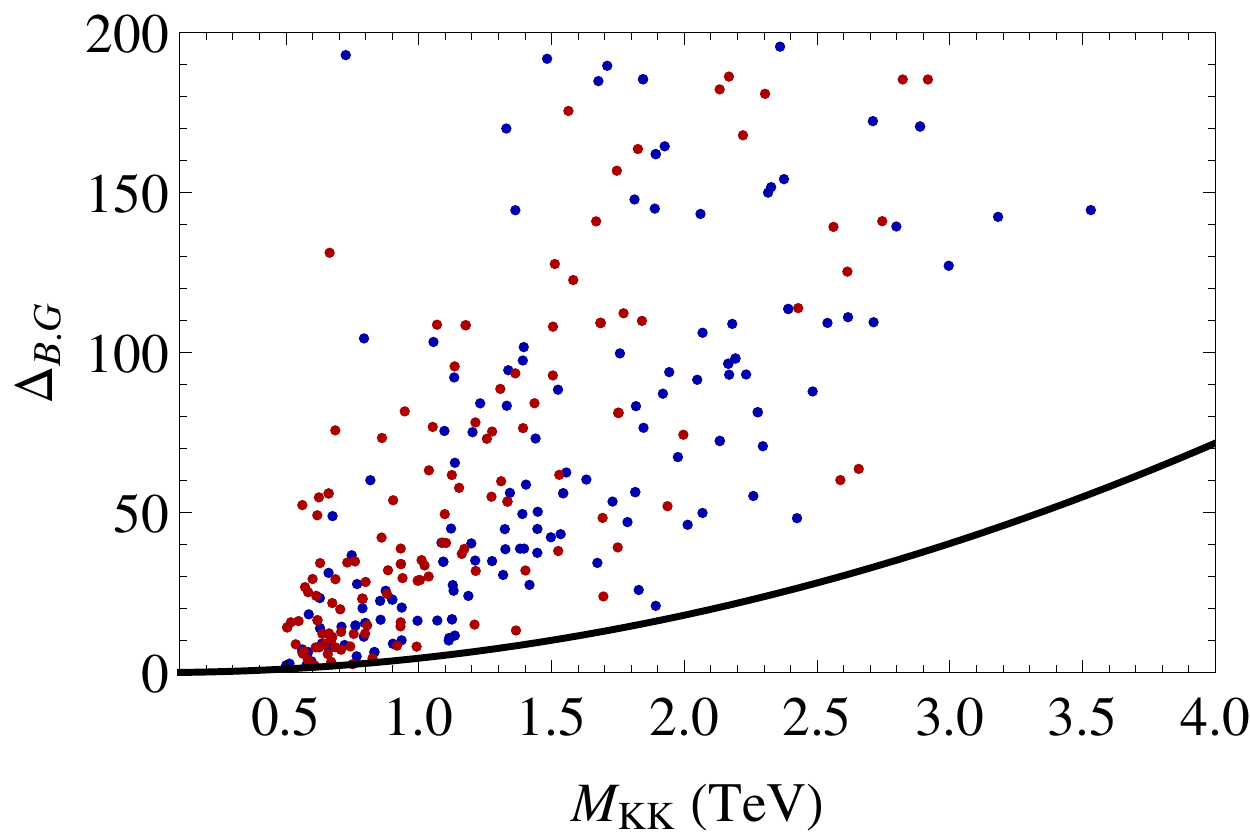}
       }
\caption{The fine-tuning as a function of $M_{\mathrm{KK}}$. Plotted in black is $\Delta_{\mathrm{B.G}}^{\mathrm{min}}=\frac{1}{s_h^2}$.   }
\label{fig:FineTuning}
\end{center}
\end{figure}

 \begin{table}[t]\scriptsize
  \centering 
 \begin{tabular}{|c|cc|cccc|}
\hline
&$\Delta_{\mathrm{B.G}}=C/s_h^2$ & $R^2$& $M_\psi=m_{+\frac{5}{3}}$&$M_\psi=m_{-\frac{4}{3}}$&$M_\psi=m_{+\frac{2}{3}}$&$M_\psi=m_{-\frac{1}{3}}$  \\ \hline
 $\theta_\psi=\theta_{\mathrm{IR}}=\zeta_{\mathrm{IR}}=0$ &$\Delta=5.4/s_h^2$ &  0.88&58\%&22\%&16\%&3\%     \\ 
$\theta_{\mathrm{IR}}=\zeta_{\mathrm{IR}}=1$, $\theta_\psi=0$ &  $\Delta=4.9/s_h^2$ & 0.68&56\%&24\%&15\%&6\%    \\ 
$\theta_{\mathrm{IR}}=3$, $\zeta_{\mathrm{IR}}=\theta_\psi=0$ &   $\Delta=5.5/s_h^2$    &    0.81 &60\%&20\%&15\%&5\%      \\ 
 $\theta_\psi=0.5$, $\theta_{\mathrm{IR}}=\zeta_{\mathrm{IR}}=0$&   $\Delta=1900/s_h^2$    &   0.04&42\%&43\%&13\%&3\%    \\
$\theta_\psi=1.5$, $\theta_{\mathrm{IR}}=\zeta_{\mathrm{IR}}=0$ &   $\Delta=1700/s_h^2$    &    0.04 &43\%&36\%&11\%&9\%\\ \hline
\end{tabular}
\caption{The second column contains the least-squares fit of $\Delta_{\mathrm{B.G}}$ to $C/s_h^2$, in addition to the associated coefficient of determination.  The third column contains the approximate percentage of the considered parameter space for which the lightest fermion resonance has a charge $+\frac{5}{3}$, $-\frac{4}{3}$, $+\frac{2}{3}$ or $-\frac{1}{3}$ .  $M_\psi=\min\{m_{+\frac{2}{3}},\;m_{-\frac{1}{3}},\;m_{+\frac{5}{3}},\;m_{-\frac{4}{3}}\}$.}\label{tab:LeastSquares }
\end{table}

The fine-tuning, as a function of $M_{\mathrm{KK}}$, has been plotted in figure \ref{fig:FineTuning}. It is generally anticipated that the minimal fine-tuning should be of the order\cite{Barbieri:2007bh, Panico:2012uw, Bellazzini:2014yua}
 \begin{displaymath}
\Delta_{\mathrm{B.G}}^{\mathrm{min}}\approx\frac{1}{s_h^2}=\frac{f_\pi^2}{v^2}.
\end{displaymath}
 Although it is also anticipated that, for models such as the MCHM${}_5$, the actual tuning will be greater than this, but still proportional to $\frac{1}{s_h^2}$, with the constant of proportionality being determined by the coefficients of the Higgs potential. In table \ref{tab:LeastSquares } we check this explicitly by making a least squares fit to $\Delta_{\mathrm{B.G}}=\frac{C}{s_h^2}$.
 
Here we are considering anarchic fermion boundary mass terms, bulk mass terms and kinetic terms. This leads to significant variation in the value of $C$. Having said that, with no kinetic terms, one finds that a reasonable fit is $\Delta_{\mathrm{B.G}}\approx\frac{5}{s_h^2}$ (which accounts for approximately 80\% of the variation). This is in partial agreement with \cite{Carena:2014ria}, who considered a two-site model with $0.1\lesssim s_h^2\lesssim 0.5$ and found $5\lesssim \Delta_{\mathrm{B.G}} \lesssim 40$. While \cite{Panico:2012uw}, considered $s_h^2=0.1$ and found the fine-tuning varying around $\Delta_{\mathrm{B.G}}\approx  40-50$.

On the other hand, when anarchic fermion kinetic terms are also included, it is no longer possible to make a meaningful fit. Nevertheless, it is found that increasing the number of anarchic input parameters, increases the chances that there will be a signifiant sensitivity to one of these parameters and so the fine-tuning typically increases. It could be argued that the fine-tuning parameter should be normalised to account for the number of input parameters. Although here we do not do this, so as to allow for comparison with other work.

 \subsection{The Mass of the Lightest Fermion Resonance}
 
 \begin{table}[h]\scriptsize
  \centering 
 \begin{tabular}{|c|ccc|ccc|ccc|}
\hline
 & \multicolumn{3}{c|}{$M_\psi>0.5$ TeV} & \multicolumn{3}{c|}{$M_\psi>1$ TeV} & \multicolumn{3}{c|}{$M_\psi>1.5$ TeV} \\
 & $\Delta<50$&   $\Delta<100$ &$\Delta<200$ &$\Delta<50$&$\Delta<100$&$\Delta<200$ &     $\Delta<50$&   $\Delta<100$ &   $\Delta<200$  \\ \hline
 $\theta_\psi=\theta_{\mathrm{IR}}=\zeta_{\mathrm{IR}}=0$ &  21\%  &  34\% &  49\% &$<1$\% & 4\%&15\%  & $<1$\%  & 1\%&6\%       \\ 
$\theta_{\mathrm{IR}}=\zeta_{\mathrm{IR}}=1$, $\theta_\psi=0$ &  32\% &  41\% & 53\%  &4\% & 9\% & 14\% &$<1$\%        &1\%       & 4\%      \\ 
$\theta_{\mathrm{IR}}=3$, $\zeta_{\mathrm{IR}}=\theta_\psi=0$ &   20\% & 30\%  & 43\% & 3\%&4\%& 13\% & 2\%       & 3\%  &  5\%     \\ 
 $\theta_\psi=0.5$, $\theta_{\mathrm{IR}}=\zeta_{\mathrm{IR}}=0$& 16\% & 34\%&49\% &4\% & 14\%& 23\%  &  $<1$\%  &$<1$\%       & 1\%      \\
$\theta_\psi=1.5$, $\theta_{\mathrm{IR}}=\zeta_{\mathrm{IR}}=0$ & 8\% & 19\% &27\%&1\% & 4\%& 9\%& $<1$\%       &  $<1$\%     &1\%  \\ \hline
\end{tabular}
\caption{An estimate of the percentage of the viable parameter space (i.e. gives rise to EWSB with the correct SM masses) that results in the lightest fermion resonance, $M_\psi$ and the fine tuning $\Delta=\Delta_{\mathrm{B.G}}$. This has been computed from the scan over the parameter space detailed in section \protect\ref{sec:ParamScan}, in particular with anarchic fermion boundary masses and $M_{\mathrm{KK}}\in[0.5,\;4]\; \mathrm{TeV}$.}\label{tab:PercentageWithFT}
\end{table}

  \begin{figure}[t]
\begin{center}

 \subfigure[]{%
            \label{fig:MPsivsFT}
            \includegraphics[width=0.45\textwidth]{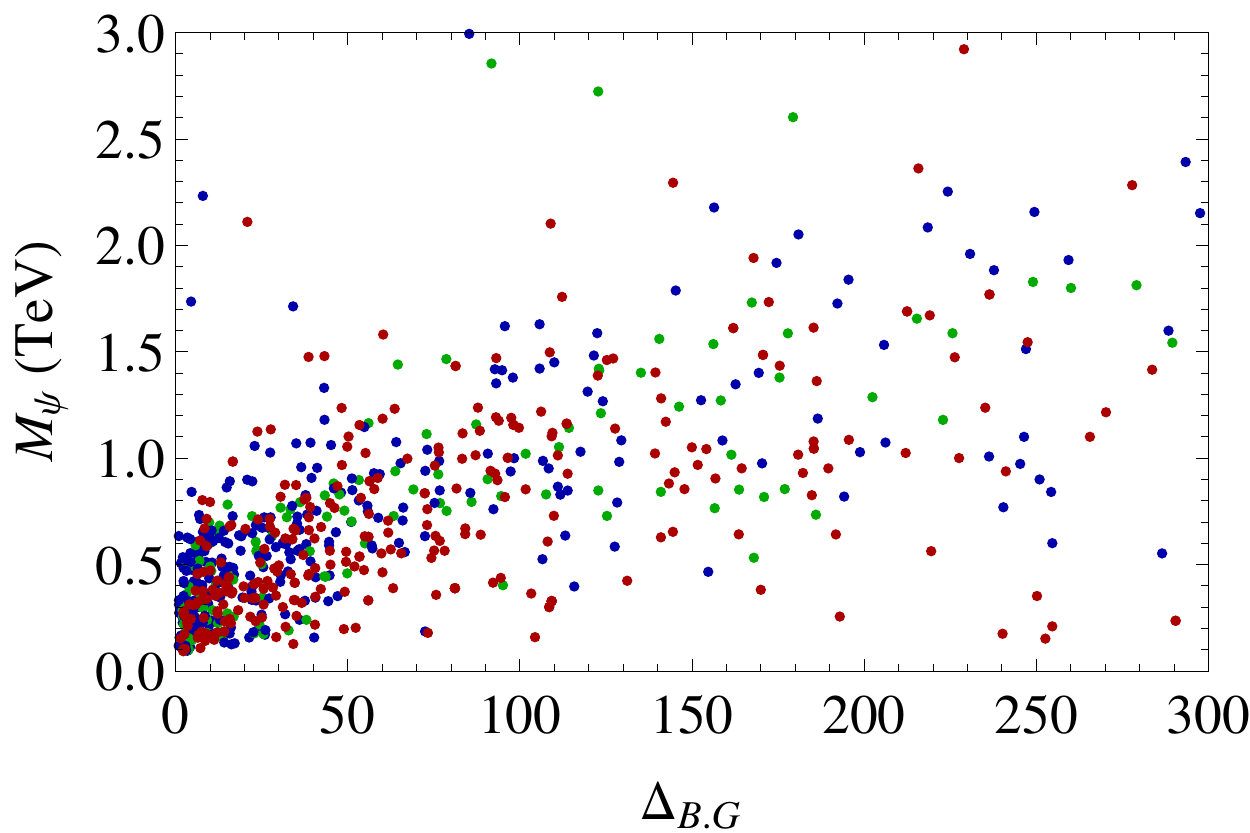}
       }
\subfigure[]{%
            \label{fig:MPsiVsMKK}
            \includegraphics[width=0.45\textwidth]{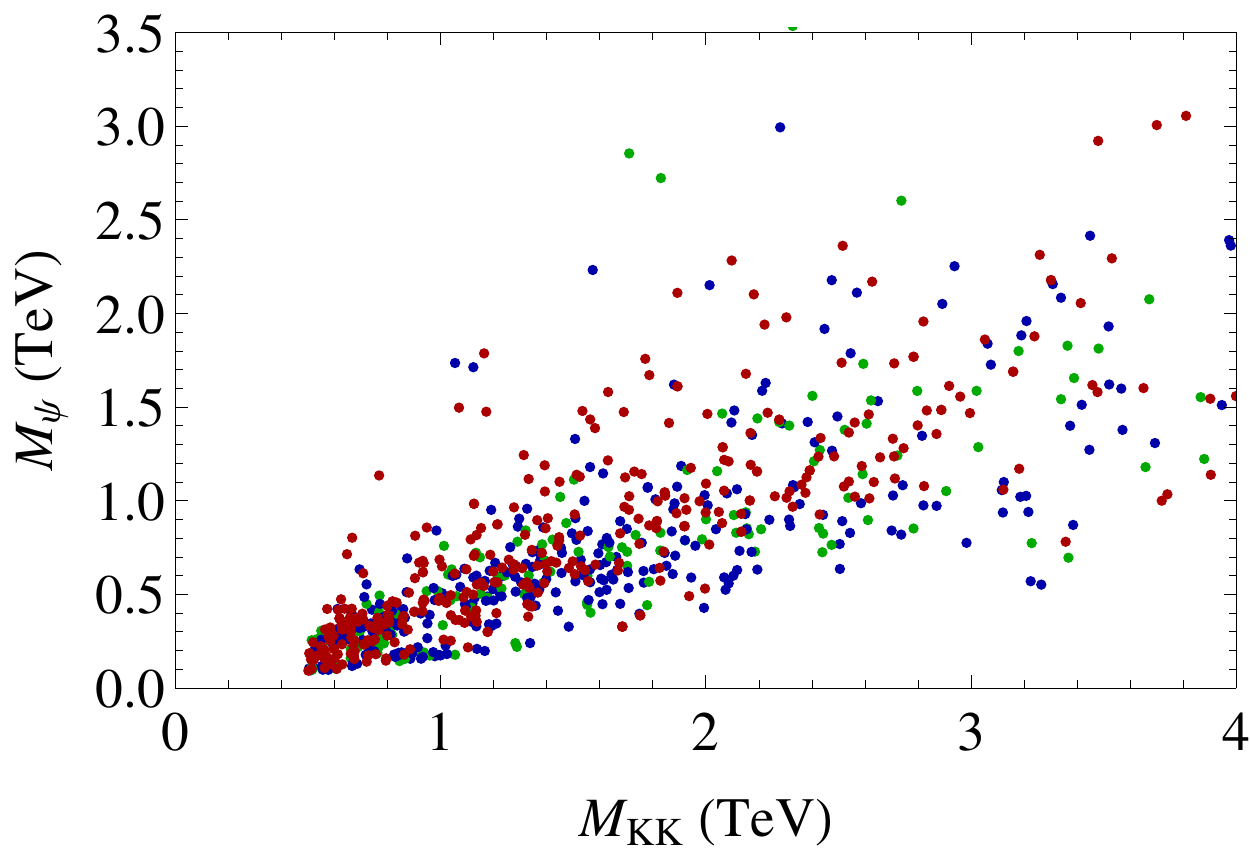}
       }\\
  \subfigure[]{%
            \label{fig:M13vsM23}
            \includegraphics[width=0.45\textwidth]{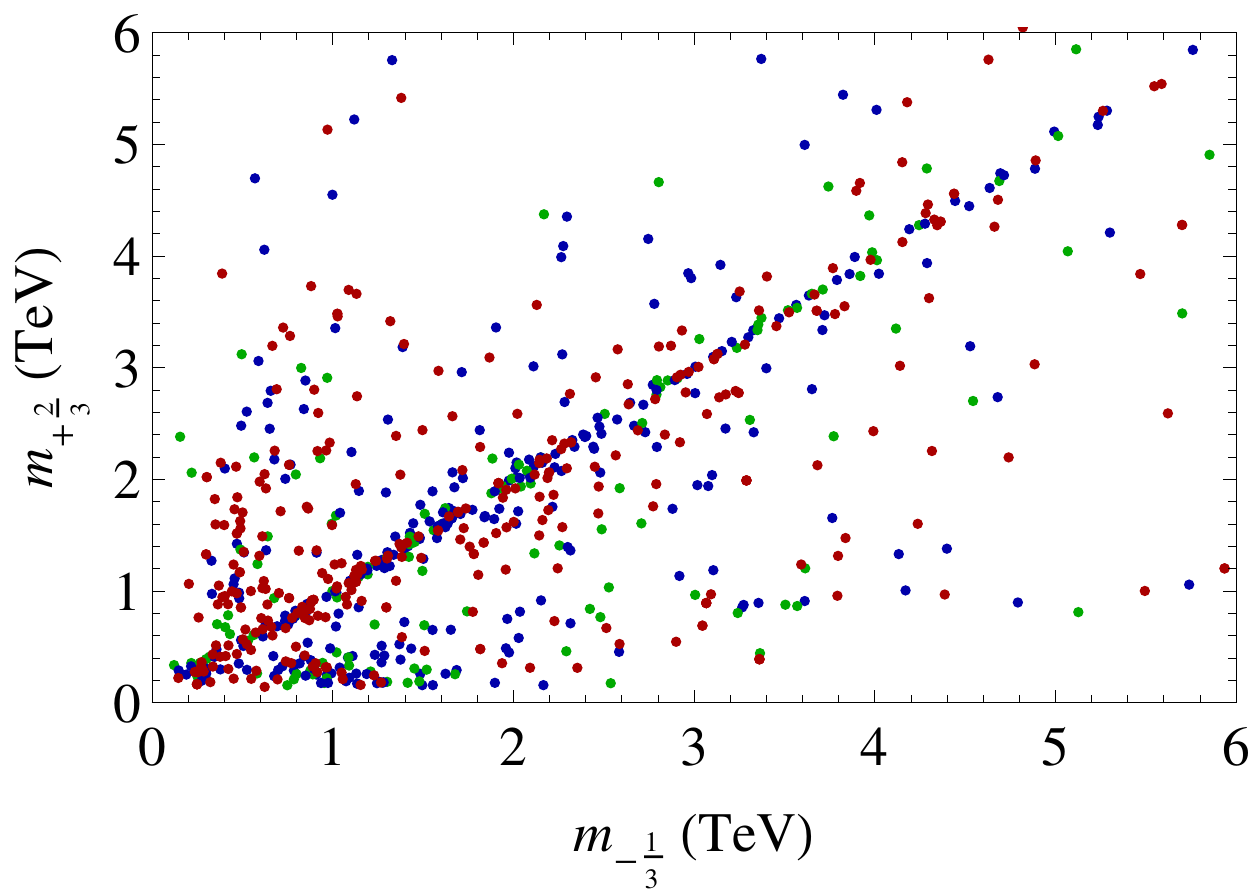}
       }
\subfigure[]{%
            \label{fig:M43vsM53}
            \includegraphics[width=0.45\textwidth]{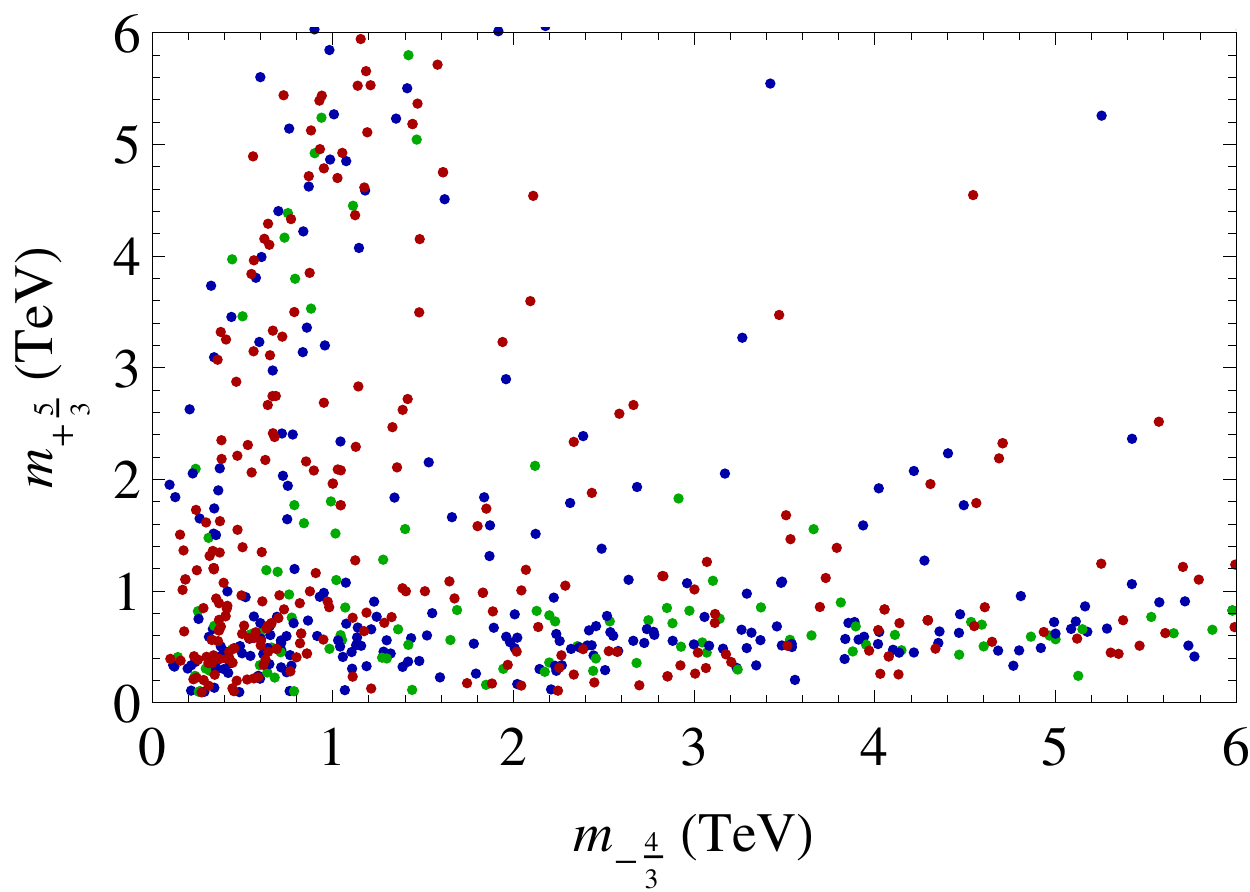}
       }      
       
\caption{Here we consider the mass of lightest fermion resonance, $M_\psi=\min\{m_{+\frac{2}{3}},\;m_{-\frac{1}{3}},\;m_{+\frac{5}{3}},\;m_{-\frac{4}{3}}\}$. The scenarios with $\theta_\psi=\theta_{\mathrm{IR}}=\zeta_{\mathrm{IR}}=0$ are plotted in green, $\theta_{\mathrm{IR}}=\zeta_{\mathrm{IR}}=1$, while $\theta_\psi=0$ and $\theta_{\mathrm{IR}}=3$, $\zeta_{\mathrm{IR}}=\theta_\psi=0$ are plotted in blue. The scenarios $\theta_\psi=0.5$, $\theta_{\mathrm{IR}}=\zeta_{\mathrm{IR}}=0$ and $\theta_\psi=1.5$, $\theta_{\mathrm{IR}}=\zeta_{\mathrm{IR}}=0$ are plotted in red.  }
\label{fig:LightestFermionRes}
\end{center}
\end{figure}
 
An important constraint comes from direct searches for heavy fermion resonances. It has been pointed out in a couple of papers \cite{Pomarol:2012qf, Matsedonskyi:2012ym}, the MCHM generically requires a light fermion resonance. Here we find no exception to this and it is found that, for minimal fine-tuning, a fermion resonance is typically required with a mass $<1$ TeV. The challenge is to try to quantify this statement. Due to the size and complexity of the parameter space it is difficult to draw any generic conclusions. However, assuming that this numerical analysis offers a fair representation of the viable parameter space and adopting a frequentist approach, we can make some comments over what is statistically likely, see figure \ref{fig:LightestFermionRes} and table \ref{tab:PercentageWithFT}.  

In particular, following partly from the correlation between $c_{q_1}$ and $c_{q_2}$, there is a tendency for viable points to have the lightest resonances $m_{+\frac{2}{3}}\approx m_{-\frac{1}{3}}$, see figure \ref{fig:M13vsM23}. Likewise, we find that in 40-60\% of the viable points the lightest fermion resonance has a charge $+\frac{5}{3}$. This is promising for LHC searches, since it would primarily decay via $T_{+\frac{5}{3}}\to tW^+\to bW^+W^+$, which can be distinguished from SM same-sign dilepton backgrounds by the presence of jets \cite{Contino:2008hi, Mrazek:2009yu, Dissertori:2010ug, DeSimone:2012fs}. If one assumes SM-like couplings and 100\% branching ratios to SM particles, then the LHC has already excluded masses $m_{+\frac{5}{3}}\lesssim 800$ GeV \cite{Chatrchyan:2013wfa} and it is projected that the LHC14 with $300\; \mathrm{fb}^{-1}$ ($3000\; \mathrm{fb}^{-1}$) will be able to discover a resonance with a mass $m_{+\frac{5}{3}} \lesssim1.38$ TeV (1.55 TeV) \cite{Avetisyan:2013rca}.

On the other hand, we also find that for between 20-40\% of the viable parameter space, the lightest fermion resonance has a charge of $-\frac{4}{3}$. These scenarios would be significantly more challenging to observe    
at the LHC since such a resonance would primarily decay via $B_{-\frac{4}{3}}\to bW^{-}$. Hence, the analogous pair production search would suffer from a large $W+\mathrm{jets}$ background and one would need a more involved search strategy focusing on single production \cite{Alvarez:2013qwa}. For completeness, the current LHC bounds on the other resonances are $m_{-\frac{1}{3}}\gtrsim 870$ GeV \cite{Aad:2013rna} and $m_{+\frac{2}{3}}\gtrsim 650$ GeV \cite{ATLAS:2012qe, Chatrchyan:2013uxa}, again assuming SM like couplings. It is projected that the LHC14 with will be able to discover masses $m_{+\frac{2}{3}} \lesssim1.2$ TeV (1.45 TeV) and $m_{-\frac{1}{3}} \lesssim1.08$ TeV (1.33 TeV) \cite{Bhattacharya:2013iea, Varnes:2013pxa}.    

Finally we come to the central question of this paper, notably how much fine-tuning does the holographic MCHM${}_5$ require? Here, again assuming this numerical analysis represents a fair representation of the true viable parameter space, we estimate that if the LHC is able to exclude a fermion resonance with a mass below 500 GeV, then it will be excluding between 20-30\% of the parameter space with $\Delta_{\mathrm{B.G}}\lesssim 50$. However if the LHC, with future luminosity upgrades, was able to exclude a fermion resonance with a mass below 1.5 TeV, then here we estimate that $\gtrsim 95\%$ of the parameter viable space with $\Delta_{\mathrm{B.G}}\lesssim 200$ would be excluded.  

Although constraints from EWPO's indicate that $M_{\mathrm{KK}}\gtrsim 2-3$ TeV which, from figure \ref{fig:MPsiVsMKK}, implies that a fermion resonance below 500 GeV should not be expected. We have not included these constraints in the above analysis, for the reasons discussed in section \ref{sect:EWPTs}, notably that the computation of the non-oblique contribution requires the specification of the level of compositeness of the first two fermion generations.

 \section{Conclusions}
\label{sec:conclu}

Describing the Higgs boson as a pNGB offers one of the most compelling explanations as to why the Higgs is lighter than the scale of new physics. However, since the model is strongly coupled, there are severe problems with calculability. This has led to progress being made by equating the model to a 5D model of gauge-Higgs unification. Of central importance, to being able to judge the viability of the MCHM, is being able to calculate the Higgs potential and hence the Higgs mass and VEV. This potential is determined by operators that violate the $\mathrm{SO}(5)$ global symmetry. Hence, in order to calculate it, one should either specify the dynamics responsible for breaking the symmetry, or work with an effective theory and include all operators permitted by the unbroken $\mathrm{SO}(4)$ symmetry. It is this latter option that is adopted in the 5D holographic descriptions of the model and in this paper.

The $\mathrm{SO}(5)$ violating operators are realised, in the 5D description, by brane localised operators and so one of the primary objectives of this paper was to investigate the implications of such operators on the Higgs potential. It is found that the gauge contribution to the potential can be suppressed by brane localised kinetic terms, but that this suppression is correlated with an enhancement in the contribution to the S parameter. Likewise the fermion contribution can be enhanced or suppressed by brane localised kinetic terms.  

However in order to realise EWSB (\ref{ViableEWSBCons}), it is necessary to either find a model, for which either $\beta$ is naturally larger than $\alpha$, or in which the $\alpha$ gauge contribution cancels with the fermion contribution. Hence simply enhancing or suppressing the particular contribution does not significantly reduce the level of fine-tuning. In practice, we find that a moderate size suppression in the gauge contribution (as is realised with the kinetic terms $\theta_{\mathrm{IR}}=\zeta_{\mathrm{IR}}=1$) gives a small improvement in the level of fine-tuning, but increasing this suppression makes the situation slightly worse. For fermion boundary terms, allowing kinetic terms significantly increases the size of the parameter space. With no physics to guide the size of these operators, we are forced to include anarchic coefficients. This increases the volatility of the Higgs potential coefficient's and typically results in a small increase in the level of tuning.

After fitting to our numerical results we find that, while there is considerable variation in the fine tuning parameter, without brane localised fermion kinetic terms the central value is approximately $\Delta_{\mathrm{B.G}}\sim \frac{5}{s_h^2}=\frac{5f_\pi^2}{v^2}$. With the inclusion of anarchic fermion kinetic terms, the fine tuning typically increases slightly due to the increased volatility in the fermion contribution. In the holographic model the value of $s_h^2$ is tightly constrained by EWPO's. In this paper we have just considered the dominant oblique contributions, since a study containing the non-oblique corrections should include all three generations as well as constraints from flavour physics. Here we leave this for future study, but note that a conservative estimate implies that $s_h^2\lesssim 0.07$, which would lead to $\Delta_{\mathrm{B.G}}^{\mathrm{min}}\gtrsim 15$ and an approximate average tuning of $\Delta_{\mathrm{B.G}}\gtrsim 70$. While these typical values, of the fine-tuning parameter, are typically much smaller than those found in supersymmetric theories, this is not really a fair comparison, since here we are considering a TeV scale effective theory without a UV completion (see \cite{Caracciolo:2012je, Barnard:2013zea} for attempts at a UV completion).  

The fine-tuning parameter is also correlated with the mass of the lightest fermion resonance (see figure \ref{fig:MPsivsFT}). We find that for the majority of the viable parameter space ($\sim 40-60$\%) the lightest fermion resonance has a charge of $+\frac{5}{3}$. Such a resonance could be seen in same sign dilepton + jets searches at the LHC and it is estimated that the LHC14 will probe $\gtrsim 95$\% of such a parameter space with  $\Delta_{\mathrm{B.G}}<200$. Although we also find a significant proportion of the parameter space ($\sim 20-40$\%) has a charged $-\frac{4}{3}$ fermion as the lightest resonance. These points would be significantly more difficult to observe at the LHC.   

So in conclusion, while the MCHM${}_5$ with minimal fine-tuning requires a fermion resonance with a mass $\lesssim500$ GeV, pre-LHC tensions with EWPO's should have indicated it was unlikely that such a particle existed. In other words,  comparing figures \ref{fig:Sparameter} and \ref{fig:MPsiVsMKK}, it is unsurprising that the LHC has not seen a new fermion with a mass $\lesssim 800$ GeV. Having said that the LHC14 will make significant inroads into the viable parameter space. It is anticipated that it will exclude a significant fraction, but not all, of the parameter space with $\Delta_{\mathrm{B.G}}<200$. 

\section{Acknowledgements} 
I would epsecially like to thank David Straub and Jose Zurita for collaboration in the early stages of this project, as well as many fruitful conversations. I am also very grateful to Eduardo Pont\'on, Marco Serone, Matthias Neubert and Lucca Vecchi for interesting and useful discussions. This work was supported by the ERC Advanced Grant EFT4LHC of the European Research Council and the Cluster of Excellence \textit{Precision Physics, Fundamental Interactions and Structure of Matter} (PRISMA-EXC 1098).

\begin{appendix}

\section{The holographic description of gauge fields}
\label{App:Gauge}
Central to the analysis of the Higgs potential, in 5D realisations of the MCHM, is the computation of the form factors appearing in (\ref{eqn:GaugeHiggsPotential} \& \ref{eqn:FermHiggsPotential}). This is most conveniently done by computing the low energy effective holographic Lagrangian. In the holographic description, rather than fixing both boundary conditions and computing the KK tower of mass eigenstates, one only fixes one of the boundary conditions and integrates out the bulk field by evaluating the field on the remaining `holographic' brane. The resulting effective Lagrangian is not given in terms of mass eigenstates, but rather a linear combination of all of the KK modes. Which brane is chosen to be the holographic brane is completely arbitrary, although in this paper we will choose the UV brane and fix our boundary conditions on the IR brane. Of course, it should be stressed that we are simply making a basis choice and the holographic Lagrangian does not rely on the AdS/CFT correspondence. Here we largely follow \cite{Panico:2007qd, Serone:2009kf}, albeit with the addition of brane localised kinetic terms.           

Let us start by considering a non-Abelian gauge field, propagating in the space (\ref{eqn:AdSMetric}) described by
\begin{equation}
\label{eqn:Gauge5DActio}
S=\int d^5x\;\sqrt{G}\left [-\frac{1}{4}F_{MN}^aF^{MN a}+\sum_{i=\scriptscriptstyle{\mathrm{IR},\;\mathrm{UV}}}\frac{\delta(r-r_i)r}{R}\left (-\frac{\tilde{\theta}_i}{4}F_{\mu\nu}^aF^{\mu\nu a}-\frac{\tilde{\zeta}_i}{2}F_{5\mu}^aF^{5\mu a}\right )\right ],
\end{equation}
where $F_{MN}^a=\partial_MA_N^a-\partial_NA_M^a+g_5f^{abc}A_M^bA_N^c$. Working in the bulk unitary gauge, such that the physical components\footnote{The caveat here is, when it exists, the zero mode of $A_5^a$, for which $\partial_rA_5^a=0$, cannot be gauged away and will give rise to a physical scalar, i.e. the Higgs in gauge-Higgs unification.} $A^a_5\rightarrow 0$, one can expand the terms quadratic in $A_\mu^a$,
\begin{align}
\label{eqn:QuadGaugeAction}
  S  &=\int d^5x\; A_\mu^a\left [\frac{R}{2r}\mathcal{P}^{\mu\nu}-\frac{\eta^{\mu\nu}}{2}\partial_r\left (\frac{R}{r}\partial_r\right )\right ]A_\nu^a   \nonumber\\
    &\hspace{1.5cm}+\sum_{i=\scriptscriptstyle{\mathrm{IR},\;\mathrm{UV}}}\delta(r-r_i)  A_\mu^a\left [\pm\eta^{\mu\nu}\frac{R}{2r}\partial_r+\frac{\tilde{\theta}_i}{2}\mathcal{P}^{\mu\nu}+\frac{\tilde{\zeta}_i}{2}\eta^{\mu\nu}\partial_r^2\right ]A_\nu^a,
\end{align}   
where $\mathcal{P}^{\mu\nu}=\eta^{\mu\nu}\Box-\partial^\mu\partial^\nu$. After Fourier transforming with respect to the four large dimensions, we disentangle the transverse and longitudinal components of $A_\mu^a$ using
\begin{equation}
\label{eqn:TransLongDef}
A^{\mu a}=\left (\eta^{\mu\nu}-\frac{p^\mu p^\nu}{p^2}\right )A_\nu^a+\frac{p^\mu p^\nu}{p^2}A_\nu^a\equiv \left(P_t^{\mu\nu}+P_l^{\mu\nu}\right )A^a_\nu= A_t^{\mu a}+A_l^{\mu a}.
\end{equation} 
This allows us to work in the Landau gauge, $A_l^{\mu a}=0$, on the holographic UV brane and equate $\eta^{\mu\nu}$ to $P_t^{\mu\nu}$. Note there is no inconsistency with working in the unitary gauge in the bulk and the Landau gauge on the brane, since one can include three independent gauge fixing terms in (\ref{eqn:Gauge5DActio}), one in the bulk and two on the branes. The equations of motion for the transverse components are then found to be
\begin{equation}
\label{eqn:GaugeEOM}
\left (\frac{r}{R}\partial_r\left (\frac{R}{r}\partial_r \right ) +p^2\right )A_t^{\mu a}=0,
\end{equation}   
where $p=\sqrt{p^2}=\sqrt{p_\mu p^\mu}$. While, variation of the IR boundary term in (\ref{eqn:QuadGaugeAction}) gives rise to two consistent BC's, that we label by parity conventions $(\pm)$ despite the fact that we consider an interval and not an orbifold,
\begin{eqnarray}
 A_t^{\mu a}\bigg{|}_{r=R^{\prime}} & =  0 &\hspace{2cm}(-),\label{eqn:GaugeNegBC}\\
\left (\frac{R}{r}\partial_r-\theta_{\mathrm{IR}}R\;p^2+\zeta_{\mathrm{IR}}R\;\partial_r^2\right )A_t^{\mu a}\bigg{|}_{r=R^{\prime}} &= 0  &\hspace{2cm}(+).\label{eqn:GaugePosBC}
\end{eqnarray}
We have also introduced the dimensionless, assumed $\mathcal{O}(1)$, coefficients $\theta_{\mathrm{IR},\mathrm{UV}}=\tilde{\theta}_{\mathrm{IR},\mathrm{UV}}/R$ and $\zeta_{\mathrm{IR},\mathrm{UV}}=\tilde{\zeta}_{\mathrm{IR},\mathrm{UV}}/R$. We now decompose the field into a four dimensional boundary field or `source field', that will act as the degree of freedom in the holographic effective action, as well as the extra dimensional form factor,
\begin{equation}
\label{eqn:GaugeDecomp}
A_t^{\mu a}(p^\mu,r)=G_A^a(p^\mu,r)\;\tilde{A}^{\mu a}(p^\mu,R).
\end{equation} 
It is then straightforward to solve (\ref{eqn:GaugeEOM}), with the $(-)$ BC (\ref{eqn:GaugeNegBC}), to find
\begin{equation}
\label{eqn:GANeg}
G_A^{a(-)}(p^\mu,r)=\frac{1}{\sqrt{1-\zeta_{\mathrm{UV}}}}\frac{r\left (\mathbf{Y}_1(pR^{\prime})\mathbf{J}_1(pr)-\mathbf{J}_1(pR^{\prime})\mathbf{Y}_1(pr)\right )}{R\left(\mathbf{Y}_1(pR^{\prime})\mathbf{J}_1(pR)-\mathbf{J}_1(pR^{\prime})\mathbf{Y}_1(pR)\right )}
\end{equation}
and for the $(+)$ BC (\ref{eqn:GaugePosBC}),

\begin{equation}
\label{eqn:GAPos}
\textstyle{
G_A^{a(+)}(p^\mu,r)=\frac{r\left [\left ((1+\zeta)\mathbf{Y}_0(pR^{\prime})-(\theta+\zeta)pR^{\prime}\mathbf{Y}_1(pR^{\prime})\right )\mathbf{J}_1(pr)-\left ((1+\zeta)\mathbf{J}_0(pR^{\prime})-(\theta+\zeta)pR^{\prime}\mathbf{J}_1(pR^{\prime})\right )\mathbf{Y}_1(pr)\right ]}{\sqrt{1-\zeta_{\mathrm{UV}}}R\left [\left ((1+\zeta)\mathbf{Y}_0(pR^{\prime})-(\theta+\zeta)pR^{\prime}\mathbf{Y}_1(pR^{\prime})\right )\mathbf{J}_1(pR)-\left ((1+\zeta)\mathbf{J}_0(pR^{\prime})-(\theta+\zeta)pR^{\prime}\mathbf{J}_1(pR^{\prime})\right )\mathbf{Y}_1(pR)\right ]}},
\end{equation}
The profiles are normalised such that $G_A^{a(\pm)}(p^\mu,R)=\frac{1}{\sqrt{1-\zeta_{\mathrm{UV}}}}$. Note this normalisation factor is completely unphysical and choosing an alternative factor would be equivalent to performing a field redefinition of the 5D gauge field.     
We have also, for ease of notation, dropped the IR subscript such that $\theta\equiv\theta_{\mathrm{IR}}$ and $\zeta\equiv\zeta_{\mathrm{IR}}$.   The holographic Lagrangian is then found by simply evaluating the UV boundary term in (\ref{eqn:QuadGaugeAction}), such that
\begin{equation}
\label{eqn:HolLag.}
\mathcal{L}_{\mathrm{Hol.}}=-\frac{1}{2}\tilde{A}_\mu^a\;P_t^{\mu\nu}\Pi^{(\pm)}(p)\tilde{A}_\nu^a
\end{equation}
where
\begin{equation}
\label{eqn:GaugePiNeg}
\Pi^{(-)}(p)=p\frac{\mathbf{Y}_1(pR^{\prime})\mathbf{J}_0(pR)-\mathbf{J}_1(pR^{\prime})\mathbf{Y}_0(pR)}{\mathbf{Y}_1(pR^{\prime})\mathbf{J}_1(pR)-\mathbf{J}_1(pR^{\prime})\mathbf{Y}_1(pR)}+\frac{\theta_{\mathrm{UV}}+\zeta_{\mathrm{UV}}}{1-\zeta_{\mathrm{UV}}}R\;p^2
\end{equation}
and
\begin{align}
\Pi^{(+)}(p)&\textstyle{=p\frac{\left ((1+\zeta)\mathbf{Y}_0(pR^{\prime})-(\theta+\zeta)pR^{\prime}\mathbf{Y}_1(pR^{\prime})\right )\mathbf{J}_0(pR)-\left ((1+\zeta)\mathbf{J}_0(pR^{\prime})-(\theta+\zeta)pR^{\prime}\mathbf{J}_1(pR^{\prime})\right )\mathbf{Y}_0(pR)}{\left ((1+\zeta)\mathbf{Y}_0(pR^{\prime})-(\theta+\zeta)pR^{\prime}\mathbf{Y}_1(pR^{\prime})\right )\mathbf{J}_1(pR)-\left ((1+\zeta)\mathbf{J}_0(pR^{\prime})-(\theta+\zeta)pR^{\prime}\mathbf{J}_1(pR^{\prime})\right )\mathbf{Y}_1(pR)}}\nonumber\\
&\hspace{9cm}\textstyle{+\frac{\theta_{\mathrm{UV}}+\zeta_{\mathrm{UV}}}{1-\zeta_{\mathrm{UV}}}R\;p^2}\label{eqn:GaugePiPos}
\end{align}
Here we have imposed $(+)$ BC's on the UV brane. One can also impose Dirichlet BC's on the UV brane. In practice this is done by including a UV localised gauge mass term and sending the mass term to infinity \cite{Panico:2007qd}. This results in the holographic source field becoming completely spurious or equivalently only fields with $(+)$ BC's remain as a local symmetry. 

In order to compute the gauge contribution to the Higgs potential, it is necessary to compute the corresponding holographic Lagrangian in Euclidean space. This is simply derived by Wick rotating $p\rightarrow ip_E$ in (\ref{eqn:GaugeEOM} \& \ref{eqn:GaugePosBC}).  This results in the effective lagrangian 
\begin{equation}
\label{eqn:HolLagEuclid.}
\mathcal{L}_{\mathrm{Hol.}}=\frac{1}{2}\tilde{A}_\mu^a\;P_t^{\mu\nu}\Pi^{(\pm)}(p_E)\tilde{A}_\nu^a
\end{equation}
where
\begin{equation}
\label{eqn:GaugePiNegEuclid}
\Pi^{(-)}(p_E)=p_E\frac{\mathbf{K}_1(p_ER^{\prime})\mathbf{I}_0(p_ER)+\mathbf{I}_1(p_ER^{\prime})\mathbf{K}_0(p_ER)}{\mathbf{K}_1(p_ER^{\prime})\mathbf{I}_1(p_ER)-\mathbf{I}_1(p_ER^{\prime})\mathbf{K}_1(p_ER)}+\frac{\theta_{\mathrm{UV}}+\zeta_{\mathrm{UV}}}{1-\zeta_{\mathrm{UV}}}R\;p_E^2
\end{equation}
and
\begin{align}
\Pi^{(+)}(p_E)&\textstyle{=p_E\frac{\left ((1+\zeta)\mathbf{K}_0(p_ER^{\prime})-(\theta+\zeta)p_ER^{\prime}\mathbf{K}_1(p_ER^{\prime})\right )\mathbf{I}_0(p_ER)-\left ((1+\zeta)\mathbf{I}_0(p_ER^{\prime})+(\theta+\zeta)p_ER^{\prime}\mathbf{I}_1(p_ER^{\prime})\right )\mathbf{K}_0(p_ER)}{\left ((1+\zeta)\mathbf{K}_0(p_ER^{\prime})-(\theta+\zeta)p_ER^{\prime}\mathbf{K}_1(p_ER^{\prime})\right )\mathbf{I}_1(pR)+\left ((1+\zeta)\mathbf{I}_0(pR^{\prime})+(\theta+\zeta)p_ER^{\prime}\mathbf{I}_1(p_ER^{\prime})\right )\mathbf{K}_1(p_ER)}}\nonumber\\
&\hspace{9cm}\textstyle{+\frac{\theta_{\mathrm{UV}}+\zeta_{\mathrm{UV}}}{1-\zeta_{\mathrm{UV}}}R\;p_E^2}.\label{eqn:GaugePiPosEuclid}
\end{align}
As already mentioned, the above Lagrangians contain a linear combination of the physical mass eigenstates and it is relatively straight forward to check that the zeroes of (\ref{eqn:GaugePiNeg} \& \ref{eqn:GaugePiPos}) correspond to the physical KK mass eigenvalues, see \cite{Batell:2007jv} for more details. Hence a possible zero in the Euclidean Lagrangian would correspond to a tachyonic mass eigenstate. In this paper, we interpret this as an indication that this is not a physically viable region of the parameter space. Examining (\ref{eqn:GaugePiPosEuclid}), we find such tachyonic modes exist when
\begin{equation}
\label{eqn:TachonCond0}
(1+\zeta_{\mathrm{IR}})\mathbf{I}_0(p_ER)\approx -(\theta_{\mathrm{IR}}+\zeta_{\mathrm{IR}})p_ER^{\prime}\mathbf{I}_1(p_ER),\quad\Rightarrow\; p_ER^{\prime}>0.
\end{equation}   
Noting that for all $x$, $\mathbf{I}_1(x)/\mathbf{I}_0(x)>0$, this then implies tachyonic modes exist when
\begin{equation}
\label{eqn:TachonCond}
\frac{\theta_{\mathrm{IR}}+\zeta_{\mathrm{IR}}}{1+\zeta_{\mathrm{IR}}}<0.
\end{equation}
 Another possible zero can occur in (\ref{eqn:HolLagEuclid.}) when there is a change of sign between the low and high momentum values of $\Pi^{(\pm)}(p_E)$. By noting that for high momenta $\frac{1}{R^{\prime}}\ll p_E\leqslant \frac{1}{R}$ one finds, 
 \begin{equation}
\label{eqn:LargeMomentomApprox}
\Pi^{(\pm)}(p_E)\approx \frac{\theta_{\mathrm{UV}}+\zeta_{\mathrm{UV}}}{1-\zeta_{\mathrm{UV}}}Rp_E^2-R\log\left (\frac{Rp_E}{2}\right )p_E^2,
\end{equation} 
 whereas for low momenta,
 \begin{equation}
\label{eqn:zeroMomeLimNeg}
\lim_{p_E\rightarrow 0}\Pi^{(-)}(p_E)=\frac{2R}{R^{\prime 2}}
\end{equation}
 and
 \begin{equation}
\label{eqn:zeroMomeLimPos }
\lim_{p_E\rightarrow 0}\frac{\Pi^{(+)}(p_E)}{p_E^2}=R\left (\log(\Omega)+\frac{\theta_{\mathrm{UV}}+\zeta_{\mathrm{UV}}}{1-\zeta_{\mathrm{UV}}}+\frac{\theta_{\mathrm{IR}}+\zeta_{\mathrm{IR}}}{1+\zeta_{\mathrm{IR}}}\right ).
\end{equation}
Then clearly another tachyonic mode can exist when
\begin{equation}
\label{eqn:HighMomTachCond}
\frac{\theta_{\mathrm{UV}}+\zeta_{\mathrm{UV}}}{1-\zeta_{\mathrm{UV}}}<0.
\end{equation}
Although this typically occurs at momenta much higher than $M_{\mathrm{KK}}$ and hence potentially at energies  beyond the scale at which we loose perturbative control of the model.

\section{The Holographic Description of Fermions}
\label{App:Fermions}
Having derived the relevant Lagrangians for the gauge sectors we shall now move on to look at fermions. The fermion sector is certainly more involved than that of the gauge sector and so, as discussed in section \ref{sect:FermCont}, here we shall simplify the scenario by just considering IR localised operators. Although it would be not to difficult to extend this work to also include UV localised operators, this would result in much larger parameter space and more complicated expressions. This work is essentially an extension of the work found in \cite{Contino:2004vy, Serone:2009kf} and matches results found in \cite{Agashe:2004rs}. The primary difference being the inclusion of brane localised kinetic terms. Although in the paper we work with just one generation, here we work with three generations primarily to serve as a reference for future work.   

\subsection{Brane Operators Consistent with a Low Energy Chiral Theory}

Fermions in five dimensions are vector like. Hence in order to write down a four dimensional effective Lagrangian, it is necessary to decompose the field into chiral components, $\Psi=\Psi_L+\Psi_R$, such that $\Psi_{L,R}=\frac{1}{2}(1\mp\gamma_5)\Psi$ and then choose either the left or right handed field to be the holographic source field. Again this choice is arbitrary. So we will now consider a scenario with two fermions, of the same gauge representation, where for one ($\psi$) we shall take the left handed component to be the source field and for the other ($\chi$), the right handed component. IR localised operators will mix these two fermions together. In particular we shall consider the following scenario,
\begin{equation}
\label{eqn:5DFermAction}
S=\int d^5x\;\sqrt{G}\;\Bigg [\sum_{\Psi=\psi,\chi}\; \left (i\bar{\Psi}_iE_A^M\gamma^A\stackrel{\leftrightarrow}{\nabla}_M\delta^{ij} \Psi_j-M^i_\Psi\bar{\Psi}_i\delta^{ij}\Psi_j\right )+\frac{\delta(r-R^{\prime})r}{R}\mathcal{L}_{\mathrm{IR}}+h.c.\Bigg ],
\end{equation}
where $G_{MN}=E^A_M\eta_{AB}E^B_N$, $\gamma^A=\{\gamma^\mu, -i\gamma^5 \}$ and the covariant derivative, $\nabla_M=D_M+\omega_M$ includes the spin connection $\omega_M=\{\frac{1}{2r}\gamma_\mu\gamma_5,0 \}$, while $\stackrel{\leftrightarrow}{\nabla}_M=\frac{1}{2}(\overrightarrow{\nabla}_M-\overleftarrow{\nabla}_M)$.  We shall also follow convention and define the dimensionless $\mathcal{O}(1)$ bulk mass parameters $c^i_\Psi=M^i_\Psi R$.  The indices $i$ and $j$ are flavour indices. It should be noted that one can, without loss of generality, rotate into a flavour basis in which the bulk operators are flavour diagonal and hence the only source of flavour and CP violation are brane localised operators \cite{Casagrande:2008hr}. After again Fourier transforming with respect to $x^\mu$ and making the field redefinition $\Psi\rightarrow \frac{r^2}{R^2}\Psi$, the resulting bulk equations of motion are found to be
\begin{equation}
\label{eqn:FermEOM}
\left(\partial_r \pm \frac{c^i_\Psi}{r}\right )\Psi_{L,R}^i=\pm \slashed{p}\Psi^i_{R,L}\hspace{1cm}\mbox{ for  }\Psi=\psi,\chi.
\end{equation}  
Naively, one may assume that there are a large number of possible IR localised operators of the form $i\bar{\Psi^i} \gamma^M\partial_M\Psi^j$. However, in order to obtain a low energy chiral theory it is necessary to impose Dirichlet BC's on either the left or right handed field. So the consistent boundary conditions are 
\begin{align}
\label{eqn:FermIRBCs}
\Psi^i_{L,R}\bigg |_{r=R^{\prime}}=0\quad (-)\quad &\hspace{0.4cm} \Rightarrow  \quad \partial_r \Psi^i_{R,L}\bigg |_{r=R^{\prime}}=\pm \frac{c^i_\Psi}{r}\Psi^i_{R,L}\bigg |_{r=R^{\prime}}\quad (+) \nonumber\\
&\quad\mbox{and}\quad \partial_r\Psi^i_{L,R}\bigg |_{r=R^{\prime}}=\pm\slashed{p}\Psi^i_{R,L}\bigg |_{r=R^{\prime}}.
\end{align} 
This reduces the number of non-zero boundary terms, in particular:
\begin{itemize}
  \item Clearly operators of the form $m\bar{\psi}_{\scriptscriptstyle{L,R}}\chi_{\scriptscriptstyle{R,L}}$, $\zeta\bar{\psi}_{\scriptscriptstyle{L,R}}\slashed{p}\chi_{\scriptscriptstyle{L,R}}$ and $\theta\bar{\Psi}_{\scriptscriptstyle{L,R}}\slashed{p}\Psi_{\scriptscriptstyle{L,R}}$ are non-zero only when the first conjugated field has a $(+)$ BC's.   
  \item Operators of the form $\bar{\Psi}_{\scriptscriptstyle{L,R}}\partial_r\Psi_{\scriptscriptstyle{R,L}}$, where $\Psi_{\scriptscriptstyle{L,R}}$ has $(+)$ BC's, are equivalent to $\bar{\Psi}_{\scriptscriptstyle{L,R}}\slashed{p}\Psi_{\scriptscriptstyle{L,R}}$ and the coefficient in front of this operator can be absorbed into the value of $\theta$.   
  \item Likewise when $\chi_{\scriptscriptstyle{R,L}}$ has a $(-)$ BC then $\bar{\psi}_{\scriptscriptstyle{L,R}}\partial_r\chi_{\scriptscriptstyle{R,L}}=\mp\bar{\psi}_{\scriptscriptstyle{L,R}}\slashed{p}\chi_{\scriptscriptstyle{L,R}}$, while when $\chi_{\scriptscriptstyle{R,L}}$ has a $(+)$ BC then $\bar{\psi}_{\scriptscriptstyle{L,R}}\partial_r\chi_{\scriptscriptstyle{R,L}}=\pm \frac{c_\chi}{R^{\prime}}\bar{\psi}_{\scriptscriptstyle{L,R}}\chi_{\scriptscriptstyle{R,L}}$ and hence the coefficients in front of both of these operators can be absorbed into the coefficients of $\zeta$ and $m$.
\end{itemize}
Hence, including operators up to mass dimension 4,  for a chiral theory the most generic IR localised Lagrangian, in (\ref{eqn:5DFermAction}), is
\begin{equation}
\label{eqn:IRFermLag}
\mathcal{L}_{\mathrm{IR}}=i\theta_{ij}^\Psi R^{\prime}\bar{\Psi}^i\slashed{\partial}\Psi^j+i\zeta_{ij}R^{\prime}\bar{\psi}^i\slashed{\partial} \chi^j-m_{ij}\bar{\psi}^i\chi^j+h.c.
\end{equation} 
where $\theta$, $\zeta$ and $m$ will be dimensionless, complex hermitian matrices. Since the source field, on the UV boundary is dynamic, the UV boundary term does not now vanish. Hence as explained in \cite{Contino:2004vy, Serone:2009kf}, it is necessary to include a UV localised Lagrange multiplier, $\mathcal{L}_{\mathrm{UV}}\supset \frac{1}{2}\left (\bar{\psi}^i\psi^i-\bar{\chi}^i\chi^i\right )$. Likewise, with the addition of the IR boundary terms (\ref{eqn:IRFermLag}), the IR boundary terms are also not automatically vanishing, but can be made to vanish with the inclusion of a Lagrange multiplier, $\mathcal{L}_{\mathrm{IR}}\supset \frac{1}{2}\left (\bar{\chi}^i\chi^i-\bar{\psi}^i\psi^i \right )$. Now the IR boundary term vanishes when    
\begin{align}
    \psi^i_R\bigg |_{r=R^{\prime}}&=\slashed{p}\theta_{ij}^\psi R^{\prime}\psi^j_L+\slashed{p}\zeta_{ij} R^{\prime}\chi^j_L-m_{ij}\chi_R^j\bigg |_{r=R^{\prime}},  \nonumber\\
   \chi^i_L\bigg |_{r=R^{\prime}} &=-\slashed{p}\theta_{ij}^\chi R^{\prime}\chi^j_R-\slashed{p}\zeta_{ij} R^{\prime}\psi^j_R+m_{ij}\psi_L^j\bigg |_{r=R^{\prime}}.  \label{FermBC1}
\end{align}
Alternatively we could have included the Lagrange multiplier $\mathcal{L}_{\mathrm{IR}}\supset \frac{1}{2}\left (\bar{\psi}^i\psi^i-\bar{\chi}^i\chi^i\right )$, which leads to the alternative consistent BC's,
\begin{align}
    \psi^i_L\bigg |_{r=R^{\prime}}&=-\slashed{p}\theta_{ij}^\psi R^{\prime}\psi^j_R-\slashed{p}\zeta_{ij} R^{\prime}\chi^j_R+m_{ij}\chi_L^j\bigg |_{r=R^{\prime}},  \nonumber\\
   \chi^i_R\bigg |_{r=R^{\prime}} &=\slashed{p}\theta_{ij}^\chi R^{\prime}\chi^j_L+\slashed{p}\zeta_{ij} R^{\prime}\psi^j_L-m_{ij}\psi_R^j\bigg |_{r=R^{\prime}}.  \label{FermBC2}
\end{align}
Before proceeding to solve (\ref{eqn:FermEOM}), it is useful to first define the functions
\begin{align}
    G^+_p(r,c)&\equiv \sqrt{r}\left (\mathbf{Y}_{c-\frac{1}{2}}(pR^{\prime})\mathbf{J}_{c+\frac{1}{2}}(pr)-\mathbf{J}_{c-\frac{1}{2}}(pR^{\prime})\mathbf{Y}_{c+\frac{1}{2}}(pr)\right ),   \nonumber\\
     G^-_p(r,c)&\equiv \sqrt{r}\left (\mathbf{Y}_{c-\frac{1}{2}}(pR^{\prime})\mathbf{J}_{c-\frac{1}{2}}(pr)-\mathbf{J}_{c-\frac{1}{2}}(pR^{\prime})\mathbf{Y}_{c-\frac{1}{2}}(pr)\right )\label{eqn:defGpGn}, 
\end{align}
such that $G^-_p(R^{\prime},c)=0$, $\left (\partial_5\pm\frac{c}{r}\right )G^\pm_p(r,c)=\pm G^\mp_p(r,c)$ and $G^+_p(R^{\prime},c_1)=G^+_p(R^{\prime},c_2)\quad \forall\; c_1,c_2$. If we choose the holographic source fields, on the UV brane, to be $\tilde{\psi}_L(p)$ and $\tilde{\chi}_R(p)$, then the general solutions to (\ref{eqn:FermEOM}) are
\begin{align}
    \psi_L^i(p,r)&=\left (A_\psi^iG^+_p(r,c_\psi^i)+B_\psi^iG^-_p(r,-c_\psi^i)\right )\tilde{\psi}^i_L(p),   \\
    \psi_R^i(p,r)&=\left (A_\psi^iG^-_p(r,c_\psi^i)-B_\psi^iG^+_p(r,-c_\psi^i)\right )\frac{\slashed{p}}{p}\tilde{\psi}^i_L(p),   \\
   \chi_R^i(p,r)&=\left (A_\chi^iG^+_p(r,-c_\chi^i)+B_\chi^iG^-_p(r,c_\chi^i)\right )\tilde{\chi}^i_R(p),   \\
    \chi_L^i(p,r)&=\left (-A_\chi^iG^-_p(r,-c_\chi^i)+B_\chi^iG^+_p(r,c_\chi^i)\right )\frac{\slashed{p}}{p}\tilde{\chi}^i_R(p). 
\end{align}
We normalise the fields such that $ \psi_L^i(p,R)=\tilde{\psi}^i_L(p)$ and $\chi_R^i(p,R)=\tilde{\chi}^i_R(p)$, which implies
\begin{equation}
\label{eqn:Adef}
A_\psi^i=\frac{1-B_\psi^i G_p^-(R,-c_\psi^i)}{G_p^+(R,c_\psi^i)}\hspace{1cm}\mbox{and}\hspace{1cm}A_\chi^i=\frac{1-B_\chi^i G_p^-(R,c_\chi^i)}{G_p^+(R,-c_\chi^i)}.
\end{equation}
The coefficients $B_{\psi, \chi}$ are then found by imposing the IR BC's. If we first define the matrices
\begin{eqnarray}
M^{ij}_\psi &\equiv \frac{m^{ij}G_p^-(R,-c_\psi^j)}{G_p^+(R,c_\psi^j)}-\zeta^{ij}pR^{\prime}, \hspace{1cm} M^{ij}_\chi&\equiv \frac{m^{ij}G_p^-(R,c_\chi^j)}{G_p^+(R,-c_\chi^j)}+\zeta^{ij}pR^{\prime}, \label{eqn:Mfermdef1}\\
   \Theta_{\psi}^{ij} &\equiv \delta^{ij}-\frac{\theta_{\psi}^{ij}pR^{\prime}G_p^-(R,-c_\psi^j)}{G_p^+(R,c_\psi^j)},\hspace{1cm}  \Theta_{\chi}^{ij} &\equiv \delta^{ij}-\frac{\theta_{\chi}^{ij}pR^{\prime}G_p^-(R,c_\chi^j)}{G_p^+(R,-c_\chi^j)}\label{eqn:ThetaFermdef1}   
\end{eqnarray}

The BC's (\ref{FermBC1}) then imply that
\begin{align}
    B_\psi\tilde{\psi}_L&=-\frac{M_\chi^{-1}\theta^\psi pR^{\prime}+\Theta_\chi^{-1}m}{M_\chi^{-1}\Theta_\psi-\Theta_\chi^{-1}M_\psi}\frac{\tilde{\psi}_L}{G^+_p(R,c_\psi)}+\frac{M_\chi^{-1}m+\Theta_\chi^{-1}\theta^\chi pR^{\prime}}{M_\chi^{-1}\Theta_\psi-\Theta_\chi^{-1}M_\psi}\frac{\slashed{p}}{p}\frac{\tilde{\chi}_R}{G_p^+(R,-c_\chi)} ,  \\
    B_\chi\tilde{\chi}_R&=-\frac{M_\psi^{-1}\theta^\chi pR^{\prime}+\Theta_\psi^{-1}m}{M_\psi^{-1}\Theta_\chi-\Theta_\psi^{-1}M_\chi}\frac{\tilde{\chi}_R}{G_p^+(R,-c_\chi)} +\frac{M_\psi^{-1}m+\Theta_\psi^{-1}\theta^\psi pR^{\prime}}{M_\psi^{-1}\Theta_\chi-\Theta_\psi^{-1}M_\chi}\frac{\slashed{p}}{p}\frac{\tilde{\psi}_L}{G^+_p(R,c_\psi)},
\end{align}
where the division of matrices denotes multiplication on the left hand side by the inverse of the matrix in the denominator. Note that 
\begin{equation}
\label{eqn:ZerolimChiralBC1}
\lim_{p\rightarrow 0}\left (\frac{M_\chi^{-1}\theta^\psi pR^{\prime}+\Theta_\chi^{-1}m}{M_\chi^{-1}\Theta_\psi-\Theta_\chi^{-1}M_\psi}\right )=\lim_{p\rightarrow 0}\left (\frac{M_\psi^{-1}\theta^\chi pR^{\prime}+\Theta_\psi^{-1}m}{M_\psi^{-1}\Theta_\chi-\Theta_\psi^{-1}M_\chi} \right )=\lim_{p\rightarrow 0}\left (\frac{G_p^-(R,c_1)}{G_p^+(R,c_2)}\right )=0,
\end{equation}
whereas
\begin{equation}
\label{eqn:ZeroLimBC1m }
\lim_{p\rightarrow 0}\left (\frac{M_\chi^{-1}m+\Theta_\chi^{-1}\theta^\chi pR^{\prime}}{M_\chi^{-1}\Theta_\psi-\Theta_\chi^{-1}M_\psi}\right )=\lim_{p\rightarrow 0}\left (\frac{M_\psi^{-1}m+\Theta_\psi^{-1}\theta^\psi pR^{\prime}}{M_\psi^{-1}\Theta_\chi-\Theta_\psi^{-1}M_\chi}\right )=m
\end{equation}
and hence, in the zero momentum limit, the 5D field $\psi_R$ would only include the source field $\tilde{\chi}_R$. In the language of the KK basis, this is equivalent to the BC's (\ref{FermBC1}) resulting in $\psi_R$ not gaining a zero mode and so one obtains an approximately chiral theory for momenta $p\ll M_{\mathrm{KK}}$.   

 Alternatively, one can note that (\ref{eqn:Adef}) is equivalent to 
\begin{equation}
\label{eqn:Bdef}
B_\psi^i=\frac{1-A_\psi^i G_p^+(R,c_\psi^i)}{G_p^-(R,-c_\psi^i)}\hspace{1cm}\mbox{and}\hspace{1cm}B_\chi^i=\frac{1-A_\chi^i G_p^+(R,-c_\chi^i)}{G_p^-(R,c_\chi^i)},
\end{equation}
then after defining
\begin{eqnarray}
\hat{M}^{ij}_\psi &\equiv \frac{m^{ij}G_p^+(R,c_\psi^j)}{G_p^-(R,-c_\psi^j)}-\zeta^{ij}pR^{\prime}, \hspace{1cm} \hat{M}^{ij}_\chi&\equiv \frac{m^{ij}G_p^+(R,-c_\chi^j)}{G_p^-(R,c_\chi^j)}+\zeta^{ij}pR^{\prime}, \label{eqn:Mfermdef2}\\
  \hat{\Theta}_{\psi}^{ij} &\equiv \delta^{ij}-\frac{\theta_{\psi}^{ij}pR^{\prime}G_p^+(R,c_\psi^j)}{G_p^-(R,-c_\psi^j)},\hspace{1cm}  \hat{\Theta}_{\chi}^{ij} &\equiv \delta^{ij}-\frac{\theta_{\chi}^{ij}pR^{\prime}G_p^+(R,-c_\chi^j)}{G_p^-(R,c_\chi^j)}\label{eqn:ThetaFermdef2}   
\end{eqnarray}
the BC's (\ref{FermBC2}) imply
\begin{align}
    A_\psi\tilde{\psi}_L&=\frac{\hat{M}_\chi^{-1}\theta^\psi pR^{\prime}-\hat{\Theta}_\chi^{-1}m}{\hat{M}_\chi^{-1}\hat{\Theta}_\psi-\hat{\Theta}_\chi^{-1}\hat{M}_\psi}\frac{\tilde{\psi}_L}{G^-_p(R,-c_\psi)}+\frac{\hat{M}_\chi^{-1}m-\hat{\Theta}_\chi^{-1}\theta^\chi pR^{\prime}}{\hat{M}_\chi^{-1}\hat{\Theta}_\psi-\hat{\Theta}_\chi^{-1}\hat{M}_\psi}\frac{\slashed{p}}{p}\frac{\tilde{\chi}_R}{G_p^-(R,c_\chi)} ,  \\
    A_\chi\tilde{\chi}_R&=\frac{\hat{M}_\psi^{-1}\theta^\chi pR^{\prime}-\hat{\Theta}_\psi^{-1}m}{\hat{M}_\psi^{-1}\hat{\Theta}_\chi-\hat{\Theta}_\psi^{-1}\hat{M}_\chi}\frac{\tilde{\chi}_R}{G_p^-(R,c_\chi)}+\frac{\hat{M}_\psi^{-1}m-\hat{\Theta}_\psi^{-1}\theta^\psi pR^{\prime}}{\hat{M}_\psi^{-1}\hat{\Theta}_\chi-\hat{\Theta}_\psi^{-1}\hat{M}_\chi}\frac{\slashed{p}}{p}\frac{\tilde{\psi}_L}{G^-_p(R,-c_\psi)}.
\end{align}
The effective holographic Lagrangian can then found by evaluating the UV boundary Lagrangian, which for the boundary conditions (\ref{FermBC1}) gives,
\begin{align}
    \mathcal{L}_{\mathrm{Hol.}}&=\frac{1}{2}\left [\overline{\psi^i_L}\psi_R^i-\overline{\chi^i_R}\chi_L^i+h.c.\right ]_{r=R}   \nonumber\\
    &= \overline{\tilde{\psi}^i_L}\;\Pi_\psi^{ij}(p)\;\slashed{p}\tilde{\psi}^j_L+ \overline{\tilde{\chi}^i_R}\;\Pi_\chi^{ij}(p)\;\slashed{p}\tilde{\chi}^j_R-\overline{\tilde{\psi}^i_L}\;M^{ij}(p)\;\tilde{\chi}^j_R-\overline{\tilde{\chi}^i_R}\;M^{ij\;\dag}(p)\;\tilde{\psi}^j_L\label{FerHolLagBC1}
\end{align} 
where
\begin{align}
    \Pi_\psi^{ij}(p)&=\textstyle{ \frac{1}{2}\Bigg (\left (G_p^+(R,-c_\psi^i)+\frac{G_p^-(R,c_\psi^i)G_p^-(R,-c_\psi^i)}{G_p^+(R,c_\psi^i)}\right )\left (\frac{M_\chi^{-1}\theta^\psi pR^{\prime}+\Theta_\chi^{-1}m}{M_\chi^{-1}\Theta_\psi-\Theta_\chi^{-1}M_\psi}\right )^{ij} \frac{1}{G_p^+(R,c_\psi^j)}}\nonumber\\
    &\hspace{9cm}\textstyle{+\frac{G_p^-(R,c_\psi^i)}{G_p^+(R,c_\psi^i)}\Bigg )\frac{1}{p}+h.c.}\nonumber\\
    \Pi_\chi^{ij}(p)&=\textstyle{ \frac{1}{2}\Bigg (\left (G_p^+(R,c_\chi^i)+\frac{G_p^-(R,c_\chi^i)G_p^-(R,-c_\chi^i)}{G_p^+(R,-c_\chi^i)}\right )\left (\frac{M_\psi^{-1}\theta^\chi pR^{\prime}+\Theta_\psi^{-1}m}{M_\psi^{-1}\Theta_\chi-\Theta_\psi^{-1}M_\chi}\right )^{ij} \frac{1}{G_p^+(R,-c_\chi^j)}}\nonumber\\
    &\hspace{9cm}\textstyle{+\frac{G_p^-(R,-c_\chi^i)}{G_p^+(R,-c_\chi^i)}\Bigg )\frac{1}{p}+h.c.}\nonumber \\
    M^{ij}(p)&=\textstyle{\frac{1}{2}\left (G_p^+(R,-c_\psi^i)+\frac{G_p^-(R,c_\psi^i)G_p^-(R,-c_\psi^i)}{G_p^+(R,c_\psi^i)}\right ) \left (\frac{M_\chi^{-1}m+\Theta_\chi^{-1}\theta^\chi pR^{\prime}}{M_\chi^{-1}\Theta_\psi-\Theta_\chi^{-1}M_\psi}\right )^{ij}\frac{1}{G_p^+(R,-c_\chi^j)}}\nonumber\\
    &\textstyle{\hspace{0.3cm}+\frac{1}{2}\left (\left (G_p^+(R,c_\chi^i)+\frac{G_p^-(R,c_\chi^i)G_p^-(R,-c_\chi^i)}{G_p^+(R,-c_\chi^i)}\right )\left (\frac{M_\psi^{-1}m+\Theta_\psi^{-1}\theta^\psi pR^{\prime}}{M_\psi^{-1}\Theta_\chi-\Theta_\psi^{-1}M_\chi}\right )^{ij}\frac{1}{G_p^+(R,c_\psi^j)}\right )^\dag.  }\label{PiMBC1}
    \end{align}
Alternatively the BC's  (\ref{FermBC2}) lead to the holographic Lagrangian
\begin{align}
    \mathcal{L}_{\mathrm{Hol.}}&=\frac{1}{2}\left [\overline{\psi^i_L}\psi_R^i-\overline{\chi^i_R}\chi_L^i+h.c.\right ]_{r=R}   \nonumber\\
    &= \overline{\tilde{\psi}^i_L}\;\hat{\Pi}_\psi^{ij}(p)\;\slashed{p}\tilde{\psi}^j_L+ \overline{\tilde{\chi}^i_R}\;\hat{\Pi}_\chi^{ij}(p)\;\slashed{p}\tilde{\chi}^j_R-\overline{\tilde{\psi}^i_L}\;\hat{M}^{ij}(p)\;\tilde{\chi}^j_R-\overline{\tilde{\chi}^i_R}\;\hat{M}^{ij\;\dag}(p)\;\tilde{\psi}^j_L\label{FerHolLagBC2}
\end{align} 
where now
\begin{align}
    \hat{\Pi}_\psi^{ij}(p)&=\textstyle{ \frac{1}{2}\Bigg (\left (G_p^-(R,c_\psi^i)+\frac{G_p^+(R,c_\psi^i)G_p^+(R,-c_\psi^i)}{G_p^-(R,-c_\psi^i)}\right )\left (\frac{\hat{M}_\chi^{-1}\theta^\psi pR^{\prime}-\hat{\Theta}_\chi^{-1}m}{\hat{M}_\chi^{-1}\hat{\Theta}_\psi-\hat{\Theta}_\chi^{-1}\hat{M}_\psi}\right )^{ij} \frac{1}{G_p^-(R,-c_\psi^j)}}\nonumber\\
    &\hspace{8cm}\textstyle{-\frac{G_p^+(R,-c_\psi^i)}{G_p^-(R,-c_\psi^i)}\Bigg )\frac{1}{p}+h.c.} \nonumber\\
    \hat{\Pi}_\chi^{ij}(p)&=\textstyle{ \frac{1}{2}\Bigg (\left (G_p^-(R,-c_\chi^i)+\frac{G_p^+(R,c_\chi^i)G_p^+(R,-c_\chi^i)}{G_p^-(R,c_\chi^i)}\right )\left (\frac{\hat{M}_\psi^{-1}\theta^\chi pR^{\prime}-\hat{\Theta}_\psi^{-1}m}{\hat{M}_\psi^{-1}\hat{\Theta}_\chi-\hat{\Theta}_\psi^{-1}\hat{M}_\chi}\right )^{ij} \frac{1}{G_p^-(R,c_\chi^j)}}\nonumber\\
    &\hspace{8cm}\textstyle{-\frac{G_p^+(R,c_\chi^i)}{G_p^-(R,c_\chi^i)}\Bigg )\frac{1}{p}+h.c.} \nonumber\\
    \hat{M}^{ij}(p)&=\textstyle{-\frac{1}{2}\left (G_p^-(R,c_\psi^i)+\frac{G_p^+(R,c_\psi^i)G_p^+(R,-c_\psi^i)}{G_p^-(R,-c_\psi^i)}\right ) \left (\frac{\hat{M}_\chi^{-1}m-\hat{\Theta}_\chi^{-1}\theta^\chi pR^{\prime}}{\hat{M}_\chi^{-1}\hat{\Theta}_\psi-\hat{\Theta}_\chi^{-1}\hat{M}_\psi}\right )^{ij}\frac{1}{G_p^-(R,c_\chi^j)}}\nonumber\\
    &\textstyle{\hspace{-0.3cm}-\frac{1}{2}\left (\left (G_p^-(R,-c_\chi^i)+\frac{G_p^+(R,c_\chi^i)G_p^+(R,-c_\chi^i)}{G_p^-(R,c_\chi^i)}\right )\left (\frac{\hat{M}_\psi^{-1}m-\hat{\Theta}_\psi^{-1}\theta^\psi pR^{\prime}}{\hat{M}_\psi^{-1}\hat{\Theta}_\chi-\hat{\Theta}_\psi^{-1}\hat{M}_\chi}\right )^{ij}\frac{1}{G_p^-(R,-c_\psi^j)}\right )^\dag.  }\label{PiMBC2}
    \end{align}
    
 \subsection{Wick Rotation of Fermion Lagrangian}   

After performing the Wick rotation, $p\rightarrow ip_E$, the form factors can be expressed in terms of the functions analogous to (\ref{eqn:defGpGn}),
\begin{align}
    G^+_{p_E}(r,c)&\equiv \sqrt{r}\left (\mathbf{K}_{c-\frac{1}{2}}(p_ER^{\prime})\mathbf{I}_{c+\frac{1}{2}}(p_Er)-\mathbf{I}_{c-\frac{1}{2}}(p_ER^{\prime})\mathbf{K}_{c+\frac{1}{2}}(p_Er)\right ),   \nonumber\\
     G^-_{p_E}(r,c)&\equiv \sqrt{r}\left (\mathbf{K}_{c-\frac{1}{2}}(p_ER^{\prime})\mathbf{I}_{c-\frac{1}{2}}(p_Er)-\mathbf{I}_{c-\frac{1}{2}}(p_ER^{\prime})\mathbf{K}_{c-\frac{1}{2}}(p_Er)\right )\label{eqn:defWickGpGn}.
\end{align} 
We can now define the matrices (\ref{eqn:Mfermdef1}, \ref{eqn:ThetaFermdef1}, \ref{eqn:Mfermdef2}, \ref{eqn:ThetaFermdef2}) in Euclidean space as  
 \begin{eqnarray}
M^{ij}_{\psi E} &\equiv -\frac{m^{ij}G_{p_E}^-(R,-c_\psi^j)}{G_{p_E}^+(R,c_\psi^j)}-\zeta^{ij}p_ER^{\prime}, \hspace{0.8cm} M^{ij}_{\chi E}&\equiv -\frac{m^{ij}G_{p_E}^-(R,c_\chi^j)}{G_{p_E}^+(R,-c_\chi^j)}+\zeta^{ij}p_ER^{\prime},\nonumber\\
\hat{M}^{ij}_{\psi E} &\equiv \frac{m^{ij}G_{p_E}^+(R,c_\psi^j)}{G_{p_E}^-(R,-c_\psi^j)}-\zeta^{ij}p_ER^{\prime}, \hspace{0.8cm} \hat{M}^{ij}_{\chi E}&\equiv \frac{m^{ij}G_{p_E}^+(R,-c_\chi^j)}{G_{p_E}^-(R,c_\chi^j)}+\zeta^{ij}p_ER^{\prime}, \label{eqn:MfermdefWick}\\
   \Theta_{\psi E}^{ij} &\equiv \delta^{ij}-\frac{\theta_{\psi}^{ij}p_ER^{\prime}G_{p_E}^-(R,-c_\psi^j)}{G_{p_E}^+(R,c_\psi^j)},\hspace{0.8cm}  \Theta_{\chi E}^{ij} &\equiv \delta^{ij}-\frac{\theta_{\chi}^{ij}{p_E}R^{\prime}G_{p_E}^-(R,c_\chi^j)}{G_{p_E}^+(R,-c_\chi^j)},\nonumber\\
   \hat{\Theta}_{\psi E}^{ij} &\equiv \delta^{ij}+\frac{\theta_{\psi}^{ij}p_ER^{\prime}G_{p_E}^+(R,c_\psi^j)}{G_{p_E}^-(R,-c_\psi^j)},\hspace{0.8cm}  \hat{\Theta}_{\chi E}^{ij} &\equiv \delta^{ij}+\frac{\theta_{\chi}^{ij}{p_E}R^{\prime}G_{p_E}^+(R,-c_\chi^j)}{G_{p_E}^-(R,c_\chi^j)}. \label{eqn:ThetaFermdefWick}   
\end{eqnarray}
This leads to the Wick rotated holographic Lagrangian, for the boundary conditions (\ref{FermBC1}), being as in (\ref{FerHolLagBC1}), but with
\begin{align}
\label{}
    \Pi_{\psi}^{ij}(p_E)&=\textstyle{ \frac{1}{2}\Bigg (\left (G_{p_E}^+(R,-c_\psi^i)-\frac{G_{p_E}^-(R,c_\psi^i)G_{p_E}^-(R,-c_\psi^i)}{G_{p_E}^+(R,c_\psi^i)}\right )\left (\frac{M_{\chi E}^{-1}\theta^\psi p_ER^{\prime}+\Theta_{\chi E}^{-1}m}{M_{\chi E}^{-1}\Theta_{\psi E}+\Theta_{\chi E}^{-1}M_{\psi E}}\right )^{ij} \frac{1}{G_{p_E}^+(R,c_\psi^j)}}\nonumber\\
    &\hspace{8.2cm}\textstyle{-\frac{G_{p_E}^-(R,c_\psi^i)}{G_{p_E}^+(R,c_\psi^i)}\Bigg )\frac{1}{p_E}+h.c.} \label{eqn:WickPiPsiBC1}\\
    \Pi_\chi^{ij}(p_E)&=\textstyle{ \frac{1}{2}\Bigg (\left (G_{p_E}^+(R,c_\chi^i)-\frac{G_{p_E}^-(R,c_\chi^i)G_{p_E}^-(R,-c_\chi^i)}{G_{p_E}^+(R,-c_\chi^i)}\right )\left (\frac{M_{\psi E}^{-1}\theta^\chi p_ER^{\prime}+\Theta_{\psi E}^{-1}m}{M_{\psi E}^{-1}\Theta_{\chi E}+\Theta_{\psi E}^{-1}M_{\chi E}}\right )^{ij} \frac{1}{G_{p_E}^+(R,-c_\chi^j)}}\nonumber\\
    &\hspace{8.2cm}\textstyle{-\frac{G_{p_E}^-(R,-c_\chi^i)}{G_{p_E}^+(R,-c_\chi^i)}\Bigg )\frac{1}{p_E}+h.c.} \label{eqn:WickPiChiBC1}\\
    M^{ij}(p_E)&=\textstyle{\frac{1}{2}\left (G_{p_E}^+(R,-c_\psi^i)-\frac{G_{p_E}^-(R,c_\psi^i)G_{p_E}^-(R,-c_\psi^i)}{G_{p_E}^+(R,c_\psi^i)}\right ) \left (\frac{M_{\chi E}^{-1}m-\Theta_{\chi E}^{-1}\theta^\chi p_ER^{\prime}}{M_{\chi E}^{-1}\Theta_{\psi E}+\Theta_{\chi E}^{-1}M_{\psi E}}\right )^{ij}\frac{1}{G_{p_E}^+(R,-c_\chi^j)}}\nonumber\\
    &\textstyle{+\frac{1}{2}\left (\left (G_{p_E}^+(R,c_\chi^i)-\frac{G_{p_E}^-(R,c_\chi^i)G_{p_e}^-(R,-c_\chi^i)}{G_{p_E}^+(R,-c_\chi^i)}\right )\left (\frac{M_{\psi E}^{-1}m-\Theta_{\psi E}^{-1}\theta^\psi p_ER^{\prime}}{M_{\psi E}^{-1}\Theta_{\chi E}+\Theta_{\psi E}^{-1}M_{\chi E}}\right )^{ij}\frac{1}{G_{p_E}^+(R,c_\psi^j)}\right )^\dag.  }
    \end{align}
While with the boundary conditions  (\ref{FermBC2}), one finds
\begin{align}
\label{}
    \hat{\Pi}_\psi^{ij}(p_E)&=\textstyle{ -\frac{1}{2}\Bigg (\left (G_{p_E}^-(R,c_\psi^i)-\frac{G_{p_E}^+(R,c_\psi^i)G_{p_E}^+(R,-c_\psi^i)}{G_{p_E}^-(R,-c_\psi^i)}\right )\left (\frac{\hat{M}_{\chi E}^{-1}\theta^{\psi} p_ER^{\prime}+\hat{\Theta}_{\chi E}^{-1}m}{\hat{M}_{\chi E}^{-1}\hat{\Theta}_{\psi E}+\hat{\Theta}_{\chi E}^{-1}\hat{M}_{\psi E}}\right )^{ij} \frac{1}{G_{p_E}^-(R,-c_\psi^j)}}\nonumber\\
    &\hspace{8.3cm}\textstyle{+\frac{G_{p_E}^+(R,-c_\psi^i)}{G_{p_E}^-(R,-c_\psi^i)}\Bigg )\frac{1}{p_E}+h.c.} \label{eqn:WickPiPsiBC2}\\
    \hat{\Pi}_\chi^{ij}(p_E)&=\textstyle{- \frac{1}{2}\Bigg (\left (G_{p_E}^-(R,-c_\chi^i)-\frac{G_{p_E}^+(R,c_\chi^i)G_{p_E}^+(R,-c_\chi^i)}{G_{p_E}^-(R,c_\chi^i)}\right )\left (\frac{\hat{M}_{\psi E}^{-1}\theta^{\chi} p_ER^{\prime}+\hat{\Theta}_{\psi E}^{-1}m}{\hat{M}_{\psi E}^{-1}\hat{\Theta}_{\chi E}+\hat{\Theta}_{\psi E}^{-1}\hat{M}_{\chi E}}\right )^{ij} \frac{1}{G_{p_E}^-(R,c_\chi^j)}}\nonumber\\
    &\hspace{8.3cm}\textstyle{+\frac{G_{p_E}^+(R,c_\chi^i)}{G_{p_E}^-(R,c_\chi^i)}\Bigg )\frac{1}{p_E}+h.c.}\label{eqn:WickPiChiBC2} \\
    \hat{M}^{ij}(p_E)&=\textstyle{-\frac{1}{2}\left (G_{p_e}^-(R,c_\psi^i)-\frac{G_{p_E}^+(R,c_\psi^i)G_{p_E}^+(R,-c_\psi^i)}{G_{p_E}^-(R,-c_\psi^i)}\right ) \left (\frac{\hat{M}_{\chi E}^{-1}m+\hat{\Theta}_{\chi E}^{-1}\theta^\chi p_ER^{\prime}}{\hat{M}_{\chi E}^{-1}\hat{\Theta}_{\psi E}+\hat{\Theta}_{\chi E}^{-1}\hat{M}_{\psi E}}\right )^{ij}\frac{1}{G_{p_E}^-(R,c_\chi^j)}}\nonumber\\
    &\textstyle{-\frac{1}{2}\left (\left (G_{p_E}^-(R,-c_\chi^i)-\frac{G_{p_E}^+(R,c_\chi^i)G_{p_E}^+(R,-c_\chi^i)}{G_{p_E}^-(R,c_\chi^i)}\right )\left (\frac{\hat{M}_{\psi E}^{-1}m+\hat{\Theta}_{\psi E}^{-1}\theta^\psi p_ER^{\prime}}{\hat{M}_{\psi E}^{-1}\hat{\Theta}_{\chi E}+\hat{\Theta}_{\psi E}^{-1}\hat{M}_{\chi E}}\right )^{ij}\frac{1}{G_{p_E}^-(R,-c_\psi^j)}\right )^\dag.  }
    \end{align}
\subsection{The Possible Existence of Tachyonic Poles}
Once again, in order to find the viable region of parameter space we must identify the values of the parameters that result in a pole in the Wick rotated propagator, or equivalently a zero in (\ref{eqn:WickPiPsiBC1}, \ref{eqn:WickPiChiBC1}, \ref{eqn:WickPiPsiBC2}, \ref{eqn:WickPiChiBC2}). For large momenta, $p_E\gg M_{\mathrm{KK}}$, the propagators are dominated by the second $\frac{G_{p_E}^\pm(R,c)}{G_{p_E}^\mp(R,c)}$ term. It can be demonstrated, for all real values of $p_E$ and $c$, that
\begin{equation}
\label{eqn:GPGNpositivity}
\frac{G_{p_E}^\pm(R,c)}{G_{p_E}^\mp(R,c)}<0,\hspace{1cm} \frac{G_{p_E}^\pm(R,c)}{G_{p_E}^\mp(R,-c)}<0 \hspace{0.5cm}\mbox{and}\hspace{0.5cm} \frac{G_{p_E}^\pm(R,c)}{G_{p_E}^\pm(R,c)}>0.
\end{equation}   
Similarly, for all values of $p_E$ and $c$, one can show that
\begin{equation}
\label{eqn:GPGNGNpositivity}
\left (G_{p_E}^+(R,c_1)-\frac{G_{p_E}^-(R,c_1)G_{p_E}^-(R,-c_1)}{G_{p_E}^+(R,-c_1)}\right )\frac{1}{G_{p_E}^+(R,c_2)}>0
\end{equation}
and
\begin{equation}
\label{eqn:GNGPGPpositivity}
\left (G_{p_E}^-(R,c_1)-\frac{G_{p_E}^+(R,c_1)G_{p_E}^+(R,-c_1)}{G_{p_E}^-(R,-c_1)}\right )\frac{1}{G_{p_E}^-(R,c_2)}<0.
\end{equation}
Hence, for $p_E\gtrsim M_{\mathrm{KK}}$, the only way (\ref{eqn:WickPiPsiBC1}, \ref{eqn:WickPiChiBC1}, \ref{eqn:WickPiPsiBC2}, \ref{eqn:WickPiChiBC2}) can contain a zero is if one of the flavour mixing matrices contains a pole. If for simplicity we consider the one generation case, then we can analyse these functions using, that for $M_{\mathrm{KK}}\lesssim p_E\ll\frac{1}{R}$,\begin{equation}
\label{ }
\frac{G_{p_E}^\pm(R,c)}{G_{p_E}^\mp(R,-c)}\approx-\frac{\mathbf{I}_{c-\frac{1}{2}}(p_ER^{\prime})}{\mathbf{I}_{-c-\frac{1}{2}}(p_ER^{\prime})}\approx -1+\frac{2c}{c^2+c-2p_ER^{\prime}}+\mathcal{O}\left (\frac{1}{pR^{\prime}}\right ).
\end{equation}
If we now make the assumption that any possible pole exists at momenta $p_E>M_{\mathrm{KK}}$, such that  $p_ER^{\prime}\gg c$, then one can approximate
\begin{equation}
\label{eqn:FlavorMixMatrixApprox}
\frac{M_{\chi E}^{-1}\theta^\psi p_ER^{\prime}+\Theta_{\chi E}^{-1}m}{M_{\chi E}^{-1}\Theta_{\psi E}+\Theta_{\chi E}^{-1}M_{\psi E}}\approx \frac{m^2+mp_ER^{\prime}+p_ER^{\prime}\theta^\psi+p_E^2R^{\prime 2}\theta^\psi\theta^\chi}{1+m^2-p^2_ER^{\prime 2}\zeta^2+p_ER^{\prime}\theta^\chi+p_ER^{\prime}\theta^\psi+p_E^2R^{\prime 2}\theta^\chi \theta^\psi},
\end{equation}
and likewise for the other matrices in (\ref{eqn:WickPiChiBC1}, \ref{eqn:WickPiPsiBC2}, \ref{eqn:WickPiChiBC2}). This function contains a pole, and hence the scenario contains a tachyonic mode, when
\begin{equation}
\label{FermionTacyonConstraint}
4\zeta^2+4m^2\zeta^2+\theta^{\chi 2}+\theta^{\psi 2}-2\theta^\psi\theta^\chi-4m^2\theta^\psi\theta^\chi>0.
\end{equation}
Interestingly, one can have both positively and negatively valued kinetic terms while still avoiding an unphysical tachyon. It is more challenging to rigorously verify the absence of tachyonic modes in the low energy region since, for example, one must take into account how composite the fermions are. Likewise, the extension of this constraint to three generations lies beyond the scope of this paper. Having said that, exhaustive numerical testing has failed to find parameters, that gave rise to a zero in (\ref{eqn:WickPiPsiBC1}, \ref{eqn:WickPiChiBC1}, \ref{eqn:WickPiPsiBC2}, \ref{eqn:WickPiChiBC2}) and did not satisfy the above constraint.

\section{Details of the Minimal Composite Higgs Model}
\label{App:MCHM}

In this appendix we will fill in a few of the details concerning the calculation in section \ref{sec:MCHM}. We will start by considering a 5D $\mathrm{SO}(5)\times\mathrm{U}(1)_X$ gauge symmetry
\begin{align}
\label{eqn:5dGaugeAct}
    S=&\int d^5x \;\sqrt{G}\;\Bigg [-\frac{1}{4g_5^2}F_{MN}^b F^{MN b}-\frac{1}{4g_{5,X}^2}X_{MN}X^{MN}  \nonumber\\
    & +\sum_{i=\scriptscriptstyle{\mathrm{IR},\;\mathrm{UV}}}\frac{\delta(r-r_i)r}{R}\left (-\frac{\tilde{\theta}_i}{4g_5^2}F_{\mu\nu}^bF^{\mu\nu b}-\frac{\tilde{\zeta}_i}{2g_5^2}F_{5\mu}^bF^{5\mu b}-\frac{\tilde{\theta}^{(X)}_i}{4g_{5,X}^2}X_{\mu\nu}X^{\mu\nu }-\frac{\tilde{\zeta}^{(X)}_i}{2g_{5,X}^2}X_{5\mu}X^{5\mu}\right )\Bigg ], 
\end{align}     
propagating in a slice of AdS${}_5$ described by (\ref{eqn:AdSMetric}). We work in a basis where the generators of $\mathrm{SO}(5)$, running over $b=1,\dots, 10$, are given by
\begin{align}
\label{}
    (T_L^a)_{\alpha\beta}&=-\frac{i}{2} \left [ \frac{1}{2}\epsilon^{abc}\left (\delta^b_\alpha\delta^c_\beta-\delta^b_\beta\delta^c_\alpha\right )+\left(\delta^a_\alpha\delta^4_\beta-\delta^a_\beta\delta^4_\alpha\right )\right ]  \nonumber\\
   (T_R^a)_{\alpha\beta}&=-\frac{i}{2} \left [ \frac{1}{2}\epsilon^{abc}\left (\delta^b_\alpha\delta^c_\beta-\delta^b_\beta\delta^c_\alpha\right )-\left(\delta^a_\alpha\delta^4_\beta-\delta^a_\beta\delta^4_\alpha\right )\right ]  \nonumber\\
    (T_C^{\hat{a}})_{\alpha\beta}&=-\frac{i}{\sqrt{2}}\left [\delta^{\hat{a}}_\alpha\delta^5_\beta-\delta^{\hat{a}}_\beta\delta^5_\alpha\right ]\label{eqn:SO5generators}
     \end{align}
where $a=1\dots 3$, $\hat{a}=1\dots 4$ and $\alpha,\beta=1\dots 5$. One then breaks the $\mathrm{SO}(5)$ to $\mathrm{SO}(4)$ by imposing `twisted' boundary condition's (BC's) \cite{Hosotani:1983xw, Hosotani:1983vn, Manton:1979kb, RandjbarDaemi:1982hi, Haba:2004qf, Hosotani:2005nz, Sakamura:2007qz,  Hosotani:2008tx} on the IR brane, in which the gauge fields in the direction of  $T_C^{\hat{a}}$ have Dirichlet BC's, while those in the direction of  $T_{L,R}^a$ have $(+)$ BC's, as in (\ref{eqn:GaugeNegBC} \& \ref{eqn:GaugePosBC}). This results in the Higgs, a complex bi-doublet under $\mathrm{SO}(4)\cong\mathrm{SU}(2)_L\times \mathrm{SU}(2)_R$, emerging as the Goldstone boson of the broken $T_C^{\hat{a}}$ generators, see (\ref{SigmaField}).

Upon EWSB, the Higgs then gains a non-zero VEV in the direction of $T_C^{\hat{4}}$, i.e. $\langle h^{\hat{a}}\rangle=(0,0,0,h)$. After expanding into the holographic basis, as described in appendix \ref{App:Gauge}, the $\mathrm{SO}(5)$ fields, $A_\mu^b=\{A_{L\mu}^a,A_{R\mu}^a,A_{C\mu}^{\hat{a}}\}$, will be related to the $\mathrm{SO}(4)$ fields, $C_{L,R\;\mu}^{\{a,a\}}$, by the transformation $A_\mu=U^\dag C_\mu U$, where   
 \begin{equation}
\label{Udef}
U=\exp \left  [i\frac{\sqrt{2}T_C^{\hat{a}}h^{\hat{a}}}{f_\pi}\right ]=\left(\begin{array}{ccccc}1 & 0 & 0 & 0 & 0 \\0 & 1 & 0 & 0 & 0 \\0 & 0 & 1 & 0 & 0 \\0 & 0 & 0 & c_h & s_h \\0 & 0 & 0 & -s_h & c_h\end{array}\right),
\end{equation} 
with $c_h=\cos (h / f_\pi)$ and $s_h=\sqrt{1-c_h^2}$. To be explicit, this results in the holographic source fields satisfying the relations
\begin{align}
\tilde{A}_{L\mu}^a & = c_{h/2}^2 \tilde{C}_{L\mu}^a+ s_{h/2}^2 \tilde{C}_{R\mu}^a \nonumber\\
\tilde{A}_{R\mu}^a & = s_{h/2}^2 \tilde{C}_{L\mu}^a+ c_{h/2}^2 \tilde{C}_{R\mu}^a \nonumber\\
\tilde{A}_{C\mu}^a & = \frac{s_h}{\sqrt{2}} \tilde{C}_{L\mu}^a- \frac{s_h}{\sqrt{2}} \tilde{C}_{R\mu}^a \nonumber\\
\tilde{A}_{C\mu}^4 & =0,
\end{align}
where $c_{h/2}=\cos \left (h/2f_\pi\right )$. There is then a further mixing, as one uses UV BC's to break $\mathrm{SU}(2)_R\times\mathrm{U}(1)_X\rightarrow \mathrm{U}(1)_Y$ \cite{Agashe:2003zs}, resulting in
\begin{equation}
\label{ SURU1toU1}
B_\mu  =s_x \tilde{C}_{R\;\mu}^3 +c_x \tilde{X}_\mu\hspace{1cm}\mbox{and}\hspace{1cm}Z^{\prime}_\mu= c_x \tilde{C}_{R\;\mu}^3 -s_x \tilde{X}_\mu  
\end{equation}
where
\begin{equation}
\label{def_sx}
s_x=\frac{g_{5,X}}{\sqrt{g_5^2+g_{5,X}^2}}\hspace{1.0cm}\mbox{ and }\hspace{1.0cm} c_x= \frac{g_{5}}{\sqrt{g_X^2+g_{5,X}^2}}.
\end{equation}
Finally, there is the familiar standard model mixing
\begin{equation}
\label{ }
W_\mu^{\pm}=\frac{1}{\sqrt{2}}\left (\tilde{C}_{L\;\mu}^1\mp i\tilde{C}_{L\;\mu}^2\right ),\hspace{1cm}Z_\mu=c_w\tilde{C}_{L\;\mu}^3-s_wB_\mu\hspace{0.5cm}\mbox{ and }\hspace{0.5cm} A_\mu=s_w\tilde{C}_{L\;\mu}^3+c_w B_\mu
\end{equation}
where the weak mixing angle is given by
 \begin{equation}
\label{WeakMixAng}
s_w=\frac{s_x}{\sqrt{1+s_x^2}}\hspace{0.5cm}\mbox{ and }\hspace{0.5cm}c_w=\frac{1}{\sqrt{1+s_x^2}}.
\end{equation}
After repeatedly using trigonometric relations such as  $\frac{1}{2}c_w^2+\frac{1}{2}s_w^2s_x^2+s_xc_ws_w=\frac{1}{2}\frac{1}{c_w^2}$, we arrive at the holographic Lagrangian
\begin{align}
\label{eqn:lagPostEWSB1}
    \textstyle{\mathcal{L}_{\mathrm{Hol.}}}&\textstyle{=-\frac{P_t^{\mu\nu}}{2}\;\bigg [\frac{2}{g_5^2}W_\mu^+\left (\Pi^{(+)}+\frac{1}{2}s_h^2\left (\Pi^{(-)}-\Pi^{(+)}\right )\right )W_\nu^-+A_\mu\left ( \frac{2s_w^2}{g_5^2}\Pi^{(+)}+\frac{c_w^2-s_w^2}{g_{5,X}^2}\Pi_X^{(+)}\right )A_\nu}\nonumber\\
    &\hspace{3cm}\textstyle{+Z_\mu \left (\frac{c_w^2+s_x^2s_w^2}{g_5^2}\Pi^{(+)}+\frac{c_x^2s_w^2}{g_{5,X}^2}\Pi_X^{(+)}+\frac{s_h^2}{2c_w^2g_5^2}\left (\Pi^{(-)}-\Pi^{(+)}\right )\right )Z_\nu}\nonumber\\
    &\hspace{6.5cm}\textstyle{+Z_\mu\;2c_ws_w\left (\frac{c_x^2}{g_5^2}\Pi^{(+)}-\frac{c_x^2}{g_{5,X}^2}\Pi_X^{(+)} \right )A_\nu\bigg ]}
\end{align}
In order to leave Lagrangian (\ref{eqn:lagPostEWSB1}) canonically normalised, we require that
\begin{equation}
\label{eqn:PiPlusNormDef}
\frac{\Pi^{(+)\prime}}{g_5^2}=\frac{\Pi_X^{(+)\prime}}{g_{5,X}^2}=\frac{1}{g^2},
\end{equation}
where $g$ is the dimensionless 4D coupling of the standard model and we have defined $\Pi^{\prime}\equiv\partial_{p^2}\Pi(p) |_{p=0}$. After expanding (\ref{eqn:GaugePiPos}) in $p^2$, this then implies that
\begin{eqnarray}
g_5^2 & = & g^2R\left (\log(\Omega)+\frac{\theta_{\mathrm{IR}}+\zeta_{\mathrm{IR}}}{1+\zeta_{\mathrm{IR}}}+\frac{\theta_{\mathrm{UV}}+\zeta_{\mathrm{UV}}}{1-\zeta_{\mathrm{UV}}}\right ), \\
g_{5,X}^2 & = & g^2R\left (\log(\Omega)+\frac{\theta_{\mathrm{IR}}^{(X)}+\zeta_{\mathrm{IR}}^{(X)}}{1+\zeta_{\mathrm{IR}}^{(X)}}+\frac{\theta^{(X)}_{\mathrm{UV}}+\zeta^{(X)}_{\mathrm{UV}}}{1-\zeta^{(X)}_{\mathrm{UV}}}\right ).
\end{eqnarray}
Further still we can also expand (\ref{eqn:GaugePiNeg}),
\begin{equation}
\label{eqn:ApproxPiNeg}
\frac{\Pi^{(-)}(p)}{g_5^2}\approx -\frac{2RM_{\mathrm{KK}}^2}{g_5^2}+\frac{R}{g_5^2}\left (\log(\Omega)-\frac{3}{4}+\frac{\theta_{\mathrm{UV}}+\zeta_{\mathrm{UV}}}{1-\zeta_{\mathrm{UV}}}\right )p^2+\mathcal{O}(\Omega^{-2}p^2)+\mathcal{O}(p^4).
\end{equation} 
It follows from our definition of $f_\pi$ that $\Pi^{(-)}(0)/g_5^2=-f_\pi^2/2$. This then implies that the SM VEV will be $v=s_hf_\pi\approx 246\; \mathrm{GeV}$, while
\begin{equation}
\label{eqn:5Dfpi}
f_\pi^2=\frac{4M_{\mathrm{KK}}^2}{g^2\left (\log(\Omega)+\frac{\theta_{\mathrm{IR}}+\zeta_{\mathrm{IR}}}{1+\zeta_{\mathrm{IR}}}+\frac{\theta_{\mathrm{UV}}+\zeta_{\mathrm{UV}}}{1-\zeta_{\mathrm{UV}}}\right )},
\end{equation}
and also
\begin{equation}
\label{eqn:sh2App }
s_h^2=\frac{m_W^2}{M_{\mathrm{KK}}^2}\left (\log(\Omega)+\frac{\theta_{\mathrm{IR}}+\zeta_{\mathrm{IR}}}{1+\zeta_{\mathrm{IR}}}+\frac{\theta_{\mathrm{UV}}+\zeta_{\mathrm{UV}}}{1-\zeta_{\mathrm{UV}}}\right ).
\end{equation}

\subsection{Fermion Holographic Lagrangian}\label{app:Fermion}

The fermion content for the MCHM${}_5$ has been specified in section \ref{sect:FermCont}, with the IR localised operators given in (\ref{MCHM5IRLag}). Note operators of the form $i\zeta \bar{\psi}\slashed{\partial} \chi$ are not present, although they could be permitted in more complicated versions of the composite Higgs model. If we now define our source fields
\begin{equation}
\label{ }
\tilde{\xi}_{q_1}=\left[\begin{array}{c}\left (\begin{array}{c} q_{1L}^\prime \\q_L\end{array} \right ) \\u_L^\prime\end{array}\right], \hspace{0.5cm}\tilde{\xi}_{q_2}=\left[\begin{array}{c}\left (\begin{array}{c} q_{L} \\q^\prime_{2L}\end{array} \right ) \\d_L^\prime\end{array}\right], \hspace{0.5cm}\tilde{\xi}_{u}=\left[\begin{array}{c}\left (\begin{array}{c} q^u_{R} \\q^\prime_{uR}\end{array} \right ) \\u_R^\prime\end{array}\right], \hspace{0.5cm}\tilde{\xi}_{d}=\left[\begin{array}{c}\left (\begin{array}{c} q^\prime_{dR} \\q^d_{R}\end{array} \right ) \\d_R^\prime\end{array}\right],
\end{equation} 
and expand in the holographic basis, as in appendix \ref{App:Fermions}. After performing the transformation, $\tilde{\xi}^\alpha\rightarrow (\Sigma^\dag)^\alpha_\beta\psi^\beta$, such that the Lagrangian is invariant under the  $\mathrm{SO}(5)$ symmetry and not just $\mathrm{SO}(4)$ \cite{Panico:2007qd}, one obtains the holographic
Lagrangian \cite{Contino:2006qr}
\begin{align}
\label{ }
\mathcal{L}_{\mathrm{hol.}}&=\sum_{r=1,2}\overline{\psi_{q_r}^\alpha}\left (\delta^{\alpha\beta}\Pi_0^{q_r}(p)+\Sigma^\alpha\Sigma^\beta\Pi_1^{q_r}(p)\right )\slashed{p}\psi_{q_r}^\beta+\sum_{r=u,d}\overline{\psi_{r}^\alpha}\left (\delta^{\alpha\beta}\Pi_0^{r}(p)+\Sigma^\alpha\Sigma^\beta\Pi_1^{r}(p)\right )\slashed{p}\psi_{r}^\beta\nonumber \\
&+\overline{\psi_{q_1}^\alpha}\left (\delta^{\alpha\beta}M_0^{q_1}(p)+\Sigma^\alpha\Sigma^\beta M_1^{q_1}(p)\right )\psi_{u}^\beta+\overline{\psi_{q_2}^\alpha}\left (\delta^{\alpha\beta}M_0^{q_2}(p)+\Sigma^\alpha\Sigma^\beta M_1^{q_2}(p)\right )\psi_{d}^\beta+h.c.\nonumber\\
&=\bar{q}_L\left [\Pi_0^q+\frac{s_h^2}{2}\left(\Pi_1^{q1}H^cH^{c\dag}+\Pi_1^{q2}HH^{\dag} \right )\right ]\slashed{p}q_L+\sum_{r=u,d}\bar{r}_R\left [\Pi_0^r+\frac{s_h^2}{2}\Pi_1^r\right ]\slashed{p}r_R\nonumber\\
&\hspace{5.5cm}+\frac{s_hc_h}{\sqrt{2}}\left (M_1^u\bar{q}_LH^cu_R+M_1^d\bar{q}_LHd_R\right )+h.c,
\end{align} 
where
\begin{equation}
\label{ Hdef}
H=\frac{1}{h}\left (\begin{array}{c}h^1-ih^2 \\ h^3-ih^4\end{array}\right )\quad\mbox{ and }\quad H^c=\frac{1}{h}\left(\begin{array}{c}-(h^1+ih^2) \\h^3+ih^4\end{array}\right).
\end{equation}
We have also projected out the fields with Dirichlet BC's on the UV brane. The form factors are given, in terms of the expressions in (\ref{PiMBC1} \& \ref{PiMBC2}), as
\begin{align}
   \Pi_0^q(p) &=\Pi_\psi\left (p,\;c_{q_1},\;c_u,\;m_u,\;\theta_L^{q_1},\;\theta_R^{u}\right )+\Pi_\psi\left (p,\;c_{q_2},\;c_d,\;m_d,\;\theta_L^{q_2},\;\theta_R^{d}\right )   \nonumber\\
     \Pi_0^u(p) &=\hat{\Pi}_\chi\left (p,\;c_{q_1},\;c_u,\;M_u,\;\theta_R^{q_1},\;\theta_L^{u}\right )\nonumber\\
     \Pi_0^d(p) &=\hat{\Pi}_\chi\left (p,\;c_{q_2},\;c_d,\;M_d,\;\theta_R^{q_2},\;\theta_L^{d}\right )\nonumber\\
     \Pi_1^{q_1}(p) &=\hat{\Pi}_\psi\left ( p,\;c_{q_1},\;c_u,\;M_u,\;\theta_R^{q_1},\;\theta_L^{u}\right )-\Pi_\psi\left (p,\;c_{q_1},\;c_u,\;m_u,\;\theta_L^{q_1},\;\theta_R^{u}\right )\nonumber\\
     \Pi_1^{q_2}(p) &=\hat{\Pi}_\psi\left ( p,\;c_{q_2},\;c_d,\;M_d,\;\theta_R^{q_2},\;\theta_L^{d}\right )-\Pi_\psi\left (p,\;c_{q_2},\;c_d,\;m_d,\;\theta_L^{q_2},\;\theta_R^{d}\right )\nonumber\\
     \Pi_1^{u}(p) &=\Pi_\chi\left (p,\;c_{q_1},\;c_u,\;m_u,\;\theta_L^{q_1},\;\theta_R^{u}\right )-\hat{\Pi}_\chi\left (p,\;c_{q_1},\;c_u,\;M_u,\;\theta_R^{q_1},\;\theta_L^{u}\right )\nonumber\\
     \Pi_1^{d}(p) &=\Pi_\chi\left (p,\;c_{q_2},\;c_d,\;m_d,\;\theta_L^{q_2},\;\theta_R^{d}\right )-\hat{\Pi}_\chi\left (p,\;c_{q_2},\;c_d,\;M_d,\;\theta_R^{q_2},\;\theta_L^{d}\right )\nonumber\\
     M_1^{u}(p) &=M\left (p,\;c_{q_1},\;c_u,\;m_u,\;\theta_L^{q_1},\;\theta_R^{u}\right )-\hat{M}\left (  p,\;c_{q_1},\;c_u,\;M_u,\;\theta_R^{q_1},\;\theta_L^{u}\right )\nonumber\\
     M_1^{d}(p) &=M\left (p,\;c_{q_2},\;c_d,\;m_d,\;\theta_L^{q_2},\;\theta_R^{d}\right )-\hat{M}\left (  p,\;c_{q_2},\;c_d,\;M_d,\;\theta_R^{q_2},\;\theta_L^{d}\right ),\label{eqn:FermPis}
\end{align}
where the order of the input parameters are $\Pi_{\psi,\chi}(p,\;c_\psi,\;c_\chi,\;m,\;\theta_\psi,\;\theta_\chi)$ and $M(p,\;c_\psi,\;c_\chi,\;m,\;\theta_\psi,\;\theta_\chi)$, while $\zeta=0$. These expression are equally valid for three generations, although we focus on the one generation scenario.

\end{appendix}

\bibliography{Higgs_potential_5d.bib}
\bibliographystyle{utphys}

\end{document}